\documentclass{article}

\usepackage{amssymb}
\usepackage{amsmath}

\newcommand{\C}{{\mathbb C}}
\newcommand{\Cl}{{\mathbb{C}\ell}}
\newcommand{\End}{{\rm End}\ }
\newcommand{\g}{{\gamma}}
\newcommand{\G}{{\Gamma}}
\newcommand{\f}{{\phi}}

\newcommand{\R}{{\mathbb R}}

\newcommand{\vf}{{\varphi}}
\newcommand{\s}{{\sigma}}
\renewcommand{\a}{{\alpha}}
\newcommand{\one}{{1\kern-2.5pt \text{l}} }
\newcommand{\cg}{{\cal G}}
\newcommand{\cp}{{\cal P}}

\newcommand{\cll}{{\cal L}}
\newcommand{\cm}{{\cal M}}
\newcommand{\Spin}{{\rm Spin}}

\numberwithin{equation}{section}

\begin{document}

\begin{center}
{\Large FROM THE GEOMETRY OF PURE SPINORS \\
WITH THEIR DIVISION ALGEBRAS\\[0.45em]
TO FERMION'S PHYSICS}\\
\vspace{1cm}
Paolo Budinich\\
International School for Advanced Studies, Trieste, Italy
E-mail: milazzi@ictp.trieste.it
\end{center}
\vspace{1cm}

\centerline{\bf Abstract}
\medskip

The Cartan's equations defining simple spinors (renamed pure by C.
Chevalley) are interpreted as equations of motion in momentum
spaces, in a constructive approach in which at each step the
dimensions of spinor space are doubled while those of momentum
space increased by two. The construction is possible only in the
frame of the geometry of simple or pure spinors, which imposes
contraint equations on spinors with more than four components, and
then momentum spaces result compact, isomorphic to
invariant-mass-spheres imbedded in each other, since the
signatures appear to be unambiguously defined and result steadily
lorentzian; starting from dimension four (Minkowski) up to
dimension ten with Clifford algebra $\Cl (1,9)$, where the
construction naturally ends. The equations of motion met in the
construction are most of those traditionally postulated ad hoc:
from Weyl equations for neutrinos (and Maxwell's) to Majorana
ones, to those for the electroweak model and for the nucleons
interacting with the pseudoscalar pion, up to those for the 3
baryon-lepton families, steadily progressing from the description
of lower energy phenomena to that of higher ones.

The $3$ division algebras: complex numbers, quaternions and
octonions appear to be strictly correlated with this spinor
geometry, from which they appear to gradually emerge in the
construction, where they play a basic role for the physical
interpretation: at the third step complex numbers generate $U(1)$,
possible origin of the electric charge and of the existence of
charged -- neutral fermion pairs, explaining also the opposite
charges of proton-electron. Another $U(1)$ appears to generate the
strong charge at the fourth step. Quaternions generate the $SU(2)$
internal symmetry of isospin and the $SU(2)_L$ one, of the
electroweak model; they are also at the origin of 3 families; in
number equal to that of quaternion units. At the fifth and last
step octonions generate the $SU(3)$ internal symmetry of flavour,
with $SU(2)$ isospin subgroup and the one of color, correlated
with $SU(2)_L$ of the electroweak model. These 3 division algebras
seem then to be at the origin of charges, families and of the
groups of the Standard model.

In this approach there seems to be no need of higher dimensional
$(> 4)$ space-time, here generated merely by Poincar\'e translations, and
dimensional reduction from $\Cl (1,9)$ to $\Cl (1,3)$ is
equivalent to decoupling of the equations of motion.

This spinor-geometrical approach is compatible with
that based on strings, since these may be expressed bilinearly (as
integrals) in terms of Majorana-Weyl simple or pure spinors which are
admitted by $\Cl (1,9)=R(32)$.

\newpage

\section{INTRODUCTION}

The discovery of fermion - and boson - multiplets which can be
labelled according to the representations of certain internal
symmetry groups $(SU(2); SU(3))$ brought to the conjecture of the
existence of a high dimensional space-time (dim. 10) in which the
ordinary one should be imbedded, and the mentioned groups are then
interpreted as rotation groups (covering of) in the extra
dimensions $(>4)$. These are then eliminated through ``dimensional
reduction''; that is by supposing they characterize compact
manifolds of very small and unobservable size.

Here we propose a more conservative and economical approach.
Reminding that, notoriously, bosons may be bilinearly expressed in
terms of fermions, we will start by considering only fermion
multiplets, which, as well known, may be represented by spinors.

The geometry of spinors was discovered by \'E. Cartan~\cite{one}
who especially stressed the great mathematical elegance of the
geometry of those spinors which he named ``simple'', subsequently
renamed ``pure'' by C. Chevalley~\cite{two}. Our proposal is to
adopt it in the study of fermion multiplets. In this
spinor-geometrical approach ordinary euclidean vector spaces will
first appear in the Cartan's equations defining spinors, which
will be immediately interpretable as equations of motion for the
fermions in momentum spaces (or in ordinary Minkowski space-time,
provided the first four components of momenta $P_\mu$ are
interpreted as generators of Poincar\'e translations). The vectors
of such spaces present some attractive properties. First they are
null, and their directions define compact manifolds. Furthermore,
as shown by Cartan, they may be bilinearly expressed in terms of
simple or pure spinors, and, if these represent the fermions, the
above equations may become identities in spinor spaces.

These will be the only ingredients for our spinor-geometrical
approach which will merely consist in starting from the simplest
non trivial two-component Dirac spinors associated with the
Clifford algebra $\Cl (2)$, equivalent to the familiar Pauli
$\Cl_0(3)$-spinors, and in constructing the fermion multiplets by
summing them directly, while building up bilinearly the
corresponding vector spaces and equations of motions in
momentum-space, step by step, which will be only possible however,
in the frame of simple or pure spinor geometry with the use of two
Propositions. At each step the dimensions of spinor space will be
doubled while those of momentum space will be increased by two. In
this construction the signature will result steadily lorentzian
while the null momenta will define spheres with radii of the
dimension of an energy, or invariant mass. Therefore in the
construction the characteristic energies of the phenomena
represented will be steadily increasing; as increasing will be the
radii of the spheres, and then the characteristic invariant
masses.

It is remarkable that in this construction, up to 32 component
spinors, (where the construction naturally ends because of the
known periodicity theorem on Clifford algebras), one naturally
finds, besides Maxwell's equations, most of those
for fermions (otherwise, historically, defined ad hoc) in momentum
spaces up to dimension $10$, associated with the Clifford algebra
$\Cl (1,9) = R(32) = \Cl (9,1)$, with signatures (including that
of Minkowski) and internal symmetry groups, unambiguously well
defined in the construction. It is also remarkable that, after
the first step which
naturally brings to Weyl equations for massless, two-component,
neutrinos, one does not arrive to Dirac's equation for massive
fermions, but rather to the equation for massive Majorana fermions
represented by four-component real spinors, associated with $\Cl
(3,1)=\R (4)$. Dirac's equation is here obtained only after the
second step, as an approximate equation, when electroweak and/or
strong interactions may be ignored.

In this construction each one of the three division algebras:
complex numbers, quaternions and octonions, seem to play a
fundamental role. In fact the former naturally appears, as soon as a
doublet of fermions is reached (at the third step) in the form of a
$U(1)$ phase symmetry for one of the two spinors, interpretable as charge
of one of the fermions of the doublet, the other being chargeless,
reminding electron-neutrino, muon-neutrino, $\dots$, proton-neutron
$\dots$ and interpretable then as electric charge. This phase symmetry,
if local, imposes gauge interactions. The next (fourth) step, bringing
to two fermion doublets, naturally produces another $U(1)$ for one of the
doublets, interpretable then as strong charge, while the quadruplet may
well represent most of properties of baryon- and lepton-doublets,
including also their similarities and their grouping in 3 families,
determined by the 3 imaginary units of the algebra of quaternions. This
algebra appears to act on fermion doublets, where it is at the origin of
isospin
$SU(2)$ symmetry of nuclear forces for the proton-neutron doublet and of
the $SU(2)_L$ of the electroweak model, for the lepton doublet. The last
and final Clifford algebra $\Cl (1,9)$ may be notoriously represented in
terms of octonions whose automorphism group $G_2$ contains a
$SU(3)$ subgroup, if one of its seven imaginary units is fixed. We
find two such $SU(3)$: one having $SU(2)$ - isospin subgroup,
interpretable then as flavour, and an orthogonal one in the
dynamical (gauge) sector of the equation interpretable as color, whose
subgroup $SU(2)_L$ appears to be correlated with the electroweak
model. These symmetries, if local determine gauge interactions
(non abelian). Therefore the groups of the standard model:
$SU(3)_C\otimes SU(2)_L\otimes U(1)$ seem naturally to originate from
these three division algebras.

In this approach the first four components of momentum
$P_1,P_2,P_3$, $P_0$ may be interpreted as Poincar\'e
generators of space-time, while the other ones
$P_5,P_6\dots P_{10}$ appear, in the equations of motion, as
interaction terms (besides the gauge ones of the dynamical sector).
Therefore, in this approach there seems to be no  need of higher dimension
space-time, while dimensional reduction will simply consist in reversing
the construction steps, by which first the terms containing $P_{10},P_9$,
then $P_8,P_7$ followed by $P_6,P_5$ will be eliminated from the
equations of motion, which means that dimensional reduction will simply
identify with decoupling of the equations of motions, and descending from
higher to lower energy phenomena.

\section{SIMPLE OR PURE SPINORS}

We will briefly summarize here some elements of spinor
geometry~\cite{one},\cite{two},\cite{three}.

Given a 2n-dimensional complex space $W= \C^{2n}$ and the
corresponding Clifford algebra $\Cl(2n)$ with generators
$\gamma_a$ obeying $\{\gamma_a,\gamma_b\}=2\delta_{ab}$, a spinor
$\phi$ is a vector of the $2^n$ dimensional representation space
$S$ of $\Cl(2n)=\End S$, defined by the Cartan's equation:
\begin{equation}
z_a\gamma^a\phi = 0;\qquad a=1,2,\dots 2n
\end{equation}
where, for $\phi\not= 0, z\in W$ may only be null and defines the Klein
quadric $Q$:
\begin{equation}
Q:\hspace{1cm} z_az^a=0\ .
\end{equation}
Define $\gamma_{2n+1}:=\gamma_1\gamma_2\dots\gamma_{2n}$,
normalized to one, as the volume element; it anticommutes with all
the $\gamma_a$ with which it generates $\Cl(2n+1)$. Weyl spinors
$\phi_+$ and $\phi_-$ are defined by
\begin{equation}
z_a\gamma^a(1\pm\gamma_{2n+1})\phi_\pm =0
\end{equation}
they are vectors of the $2^{n-1}$   -dimensional representation
space of the even subalgebra $\Cl_0(2n)$ of $\Cl(2n)$.

Given $\phi_\pm$ eq.(2.3) defines a $d$-dimensional totally null plane
$T_d(\phi_\pm )$ whose vectors are all null and mutually orthogonal.
For $d=n$; that is for maximal dimension of the totally null plane
$T_n(\phi_\pm ):=M(\phi_\pm )$, the corresponding Weyl spinor $\phi_\pm$
is named simple or pure and, as stated by Cartan, $M(\phi_\pm )$ and
$\pm\phi_\pm$ are equivalent. All Weyl
spinors are simple or pure for $n\leq 3$. For $n\geq 4$ simple or
pure Weyl spinors are subject to a number of constraint equations
which equals $1, 10, 66, 364$ for $n = 4, 5, 6, 7$ respectively (see
eq.(2.8)).

To express the vectors $z\in W$ (and tensors) bilinearly in the
terms of spinors we need to define the main automorphism $B$ of
$\Cl(2n)$ through:
\begin{equation}
B\gamma_a=\gamma^t_aB;\quad B\phi =\phi^tB\in S^*
\end{equation}
where $S^*$ is the dual spinor space of $S$,  $\gamma^t_a$ and
$\phi^t$ meaning $\gamma_a$ and $\phi$ transposed.

In the case of real, pseudoeuclidean, vector spaces, of interest for
physics, we will also need to define the conjugation operator $C$
such that
\begin{equation}
C \g_a = {\bar \g_a} C \quad {\rm and } \quad \f^c = C \bar \f
\end{equation}
where $\bar \g_a$ and $\bar \f$ mean $\g_a$ and $\f$ complex
conjugate.

Another useful definition of simple spinors may be obtained
through the formula~\cite{four}:
\begin{equation}
\f \otimes B \psi = \sum\limits_{j=0}^n F_j
\end{equation}
where $\f, \psi \in S$ are spinors of $\Cl(2n) = \End S$ and
$$
  F_j = {}_[{}\raisebox{0.5ex}{$\g_{a_1} \g_{a_2} \cdots
\g_{a_j}$}{}_] T^{a_1 a_2\dots  a_j}
\eqno(2.6')
$$
where the $\g_a$ products are antisymmetrized and $ T^{a_1 a_2
\dots a_j}$ are the components of an antisymmetric $j$-tensor of
$\C^{2n}$, which can be expressed bilinearly in terms of the
spinors $\f$ and $\psi$ as follows:
\begin{equation}
  T_{a_1 a_2 \dots a_j}= \frac{1}{2^n} \langle B \psi,
{}_[{}\raisebox{0.5ex}{$ \g_{a_1}
  \g_{a_2} \cdots \g_{a_j }$}{}_] \f \rangle .
\end{equation}
Setting now $\psi = \f$, in eq.~(2.6) we have that: $\f$ is simple or pure
if and only if
\begin{equation}
  F_0=0,\quad F_1 = 0, \quad F_2 = 0,\dots , F_{n-1} = 0
\end{equation}
while $F_n \not= 0$ and eq.~(2.6) becomes:
\begin{equation}
   \phi \otimes B \phi = F_n
\end{equation}
and the $n$-tensor $F_n$ represents the maximal totally null plane
$M(\phi )$ of $W$ equivalent, up to a sign, to the simple spinor
$\f$. Equations~(2.8) represent then the constraint equations for
a spinor $\f$ associated with $W$ to be a simple or pure.
It is easy to verify that for $n\leq 3$, $F_0,F_1,F_2$ are
identically zero for symmetry reasons, while for $n=4$, the only
constraint equation to be imposed is $F_0=0$. For $n=5,6$ the
constraint equations are $F_0=F_1=0$; and $F_0=F_1=F_2=0$ in
numbers $10$ and $66$ respectively.

The equivalence of this definition with the one deriving from
eq.~(2.1), given by Cartan, is easily obtained if we multiply
eq.~(2.6) on the left by $\g_a$ and on the right by $\g_a\f$ and
sum over $a$, obtaining
\begin{equation}
    \g_a \f \otimes B \psi \g^a \f = z_a \g^a \f
\end{equation}
where
\begin{equation*}
z_a = \frac{1}{2^n}\langle B \psi, \g_a \f \rangle
\end{equation*}
which, provided $\f$ is simple or pure , for arbitrary $\psi$, satisfy
\begin{equation*}
  z_a \g^a \f = 0
\end{equation*}
and $z_a$ are the components of a null vector of $W$, belonging to
$M(\phi )$, and we have the following proposition:\\

\noindent{\bf Proposition 1.}\ Given a complex space $W=\C^{2n}$
with its Clifford algebra $\Cl (2n)$, with generators $\gamma_a$,
let $\psi$ and $\phi$ represent two spinors of the endomorphism
spinor-space of $\Cl (2n)$ and of its even subalgebra $\Cl_0(2n)$,
respectively. Then, the vector $z\in W$, with components:
\begin{equation}
z_a =\langle B\psi ,\gamma_a\phi\rangle ;\quad a=1,2,\dots 2n
\end{equation}
is null, for arbitrary $\psi$, if and only if $\phi$ is a simple or pure
spinor.

The proof is given in reference~\cite{five}.

We have now listed the main geometrical instruments of simple or
pure  spinor geometry useful in order to proceed with the program.

We will start with the simplest non trivial case of two component
spinors, and will perform the first natural
step bringing to four component spinors, which, while
already well known~\cite{six}, is here reformulated in order to exibit
transparently some of the main features of Cartan's spinor geometry,
and the possible algebraic origin of the Minkowski signature of
space-time.

\section{THE FIRST NATURAL STEP FROM TWO TO FOUR COMPONENT
SPINORS AND THE SIGNATURE OF SPACE-TIME}
\subsection{Weyl equations}

Let us start from $W = \C^2$, with Clifford algebra $\Cl (2)$,
generators $\s_1,\s_2$ satisfying $\{\s_j,\s_k\} =2\delta_{jk}$
and volume element $\s_3=-i\s_1\s_2$. $\Cl (2)$ is simple and
isomorphic to $\Cl_0(3)=\End S$ generated by the Pauli matrices
$\s_1,\s_2,\s_3$. Its Pauli spinors $\vf =
\begin{pmatrix} \vf_0 \cr \vf_1 \end{pmatrix}
\in S$ are simple. In fact we have: $B= -i \sigma_2 =
\begin{pmatrix} 0 & -1 \cr 1 & 0
\end{pmatrix} :=\epsilon$ and $F_0=\langle \epsilon\vf,\vf \rangle \equiv
0$; eq.~(2.9) becomes:
\begin{equation}
  \begin{pmatrix} \vf_0 \vf_1 & - \vf_0 \vf_0 \cr \vf_1 \vf_1 &
  -\vf_1
  \vf_0 \end{pmatrix}\equiv \vf \otimes B \vf = z_j \s_j \equiv
  \begin{pmatrix} z_3 & z_1 -i z_2 \cr z_1+ i z_2 & -z_3 \end{pmatrix}\ .
\end{equation}
Furthermore we have:
$$
z_j=\frac{1}{2}\langle B\vf, \s_j \vf\rangle
=\frac{1}{2}\varphi^t\epsilon\s_j\varphi
\eqno(3.1')
$$
(compare the elements of the two matrices). From eqs.~(3.1)and
(3.1$'$), equation:
\begin{equation}
z_1^2 + z_2^2 + z_3^2 = 0
\end{equation}
follows and is identically satisfied as may be immediately seen
from the determinants of the matrices in eq.~(3.1). Also the
Cartan's equation:
\begin{equation}
z_j \s^j \vf =0
\end{equation}
is identically satisfied, (as may be immediately seen if we act
with the first matrix of eq.~(3.1) on the spinor $\vf =\begin{pmatrix}
\vf_0\cr \vf_1\end{pmatrix}$).

If $\psi \in S$ is another spinor we have, from eq.~(2.6):
\begin{equation}
  \begin{pmatrix}
  \vf_0 \psi_1 & - \vf_0 \psi_0 \cr \vf_1 \psi_1 & -\vf_1
  \psi_0 \end{pmatrix}\equiv \vf \otimes B \psi = z_0 + z_j \s_j \equiv
  \begin{pmatrix} z_0 +z_3 & z_1-iz_2 \\ z_1+iz_2 & z_0 - z_3
  \end{pmatrix} ,
\end{equation}
and $z_0 = \frac{1}{2}\psi^t\epsilon\vf$;\ \ \ $z_j =
\frac{1}{2}\psi^t\epsilon \s_j \vf$, deriving from it, satisfy
identically the equation (as may be immediately seen from the
determinants of the matrices):
\begin{equation}
   z_1^2 + z_2^2 + z_3^2 - z_0^2 = 0
\end{equation}
which uniquely determines the signature of Minkowski. In fact
the above may be easily restricted to the real by substituting $B\psi$ with
$B\phi^c = \phi^\dagger$ by which $z_0$, $z_j$ become $p_0$, $p_j$
real:
\begin{equation}
    p_0 = \frac{1}{2}\langle\vf^\dagger \vf\rangle ; \qquad p_j =
\frac{1}{2}\langle \vf^\dagger \s_j \vf \rangle
\end{equation}
satisfying identically to:
\begin{equation}
    p_1^2 + p_2^2 + p_3^2 - p_0^2 = 0\ ;
\end{equation}
a null or optical vector of Minkowski momentum space $P=\R^{3,1}$,
which then originates from the structure of Pauli  matrices (or
quaternion units): see the second matrix in (3.4) or, equivalently,
from the Clifford algebras isomorphism:
\begin{equation}
    \Cl(3) \simeq \Cl_0(3,1) ,
\end{equation}
after which $\vf$ may be interpreted as a simple Weyl spinor
associated with $\Cl_0(3,1) = \End S_\pm$; and, since $\Cl_0(3,1)$
is not simple, there are two of them: $\vf_+$, $\vf_-$ satisfying
the Cartan's equations:
\begin{equation}
\left( \vec p \cdot \vec \s + p_0 \right) \vf_+=0 \qquad \left( \vec p
\cdot \vec \s - p_0 \right) \vf_-=0\ .
\end{equation}

These equations may be expressed as a single equation for the four
component Dirac spinor $\psi =\vf_+ \oplus \vf_-$. In
fact indicating with
\begin{equation*}
    \g_\mu = \left\{\g_0; \g_j\right\}:=
    \left\{-i\sigma_2\otimes 1; \sigma_1\otimes \sigma_j\right\} ;\
    \ j=1,2,3
\end{equation*}
the generators of $\Cl(3,1)$ and with $\g_5 = -i\g_0 \g_1 \g_2
\g_3 = \s_3 \otimes 1$ its volume element, we may write (3.9) in
the form
$$
  p^\pm_\mu \g^\mu \left( 1 \pm \g_5\right) \psi = 0
    \eqno(3.9')
$$
where the Weyl spinors $\vf_\pm = \frac{1}{2}\left(1 \pm
\g_5\right) \psi$ are simple or pure; they are eigenspinors of $\g_5$.

The optical vectors $p^\pm_\mu$ may be expressed in the form:
\begin{equation}
p_\mu^\pm = \frac{1}{2}{\tilde \psi} \g_\mu \left(1 \pm
\g_5\right) \psi ,
\end{equation}
where ${\tilde \psi} = \psi^\dagger \g_0$; with which  eq.(3.7),
becomes an identity. Eq.(3.9$'$) is a particular case of Cartan's
eq.(2.3) defining simple or pure spinor and as such it may be
dealt with as an algebraic equation\footnote{One may define
spinors as minimal left ideals of Clifford algebras and then, in
the Witt nilpotent basis of $\Cl (3,1)$, simple or pure spinors
are products of basis elements as $\psi_0=n_1n_2$ say, with
$n_1=\frac{1}{2}(\gamma_1+i\gamma_2);
n_2=\frac{1}{2}(\gamma_3+\gamma_0)$; and eq.(3.9$'$) becomes
$(x_1n_1+x_2n_2)\psi_0=0$ satisfied for any $x_1=p_1 -ip_2$ and $
x_2=p_3-p_0$\cite{four} .}. To interpret it as an equation of
motion we have to postulate that the spinor $\psi :=\psi (p)$ is a
function of $p\in P=\R^{3,1}$ (one could conceive the Clifford
algebra $\Cl_0(3,1)$ as a fiber in $P=\R^{3,1}$) and then
eqs.(3.9$'$) represent the Weyl equations for massless neutrinos
in momentum space $P$.

But one could also think of $p_\mu$ as generators of Poincar\'e
translations through which we may notoriously generate, as a
homogeneous space, Minkowski space-time $M=\R^{3,1}$. Then if
$x\in M$, $p_\mu=i\frac{\partial}{\partial x_\mu}$ and $\psi
:=\psi (x)$ (again we could conceive $\Cl_0(3,1)$ as a fiber in
$M$) and eqs.(3.9$'$) become
\begin{equation}
i\ \frac{\partial}{\partial x_\mu}\ \gamma^\mu (1\pm\gamma_5)\psi (x)=0
\end{equation}
that is Weyl equations for massless neutrinos in space-time $M$.
These two options are both mathematically correct. In this case
they are equivalent since eq.(3.9$'$) is the Fourier transform of
eq.(3.11). Observe that in this approach space-time appears merely
as generated by Poincar\'e translations.

If $p^\pm_\mu$ given in eq.(3.10) is inserted
in the equation of motion (3.9$'$), this appears expressed in terms of
simple or pure spinors associated with
$\Cl_0 (3,1)$ (and it can also become an identity) and therefore
establishes a correlation of that geometry with the physics of massless
neutrinos (like non conservation of parity). The solution  $\psi (x)$ of
eq.(3.11) describes the space-time behaviour of a massless neutrino in
a particular phenomenon: the fact that, if inserted in (3.11), it makes
it an identity means that such description is valid for any
space and time coordinate. These properties of Cartan's eqs.(3.9$'$) and
(3.11) are general and valid also for the other equations we will deal
with in the paper.

\subsection{Maxwell's equations}

Also Maxwell's equations may be easily obtained from eq.(3.9$'$).
In fact the Weyl spinors $\vf_\pm$ are simple and therefore with them
eq.~(2.9) becomes:
\begin{equation}
  \vf_\pm \otimes B \vf_\pm = \frac{1}{2} F_\pm^{\mu\nu} \left[ \g_\mu,
  \g_\nu \right] \left( 1\pm\g_5\right)
\end{equation}
where the antisymmetric tensor $F^{\mu\nu}_\pm$ may be expressed
bilinearly in terms of spinors through eq.~(2.7):
\begin{equation}
  F_\pm^{\mu\nu} = \tilde \psi \left[ \gamma^\mu, \gamma^\nu \right]
  \left( 1\pm \gamma_5 \right) \psi \,,
\end{equation}
and $F_\pm^{\mu\nu}$, because of eq.(3.9$'$) satisfy
the homogeneous Maxwell's equations~\cite{seven},\cite{eight}:
\begin{equation}
p_\mu F^{\mu\nu}_+=0;\qquad \epsilon_{\lambda\rho\mu\nu}p^\rho
F^{\mu\nu}_-=0 .
\end{equation}

Observe that from eq.(3.13) it appears that the electromagnetic
tensor $F_{\mu\nu}$ is bilinearly expressed in terms of the Weyl
spinors $\vf_+$ and $\vf_-$ obeying the equations of motion
(3.9$'$) of massless neutrinos. This however does not imply
that, in the quantized theory, the photon must be conceived as a
bound state of neutrinos.  In fact it is known that the neutrino
theory of light, while violating both gauge invariance and
statistics, is unacceptable~\cite{nine}.

\section{IMBEDDING SPINORS AND NULL-VECTOR SPACES IN HIGHER
DIMENSIONAL ONES}

Our programme is to imbed spinor spaces in higher dimensional ones,
and this is easy with the instrument of direct sums; but we want
also to imbed the corresponding null vector spaces in higher
dimensional ones, which might appear difficult since our momentum
space vectors are bilinearly constructed from spinors
and furthermore, sums of null vectors
are not null, in general. This will be possible instead only in
the frame of simple- or pure-spinors geometry, as we will see.

Let us start by considering the isomorphisms of Clifford algebras:
\begin{equation}
\Cl(2n)\simeq \Cl_0(2n+1)
\end{equation}
both being central simple~\cite{three}, and
\begin{equation}
\Cl(2n+1)\simeq \Cl_0(2n+2)
\end{equation}
both non simple. From these we may have the following imbeddings:
\begin{equation}
\Cl(2n)\simeq \Cl_0(2n+1) \hookrightarrow \Cl(2n+1)\simeq
\Cl_0(2n+2) \hookrightarrow \Cl(2n+2) .
\end{equation}

We will indicate with $\psi_D,\psi_P$, and $\psi_W$ the spinors
associated with $\Cl(2n)$, $\Cl_0(2n+1)$ and $\Cl_0(2n)$
respectively, where $D, P$ and $W$ stand for Dirac,
Pauli and Weyl. Then to the embeddings (4.3) of Clifford algebras
there correspond the following spinor embeddings:
\begin{equation}
\psi_D\simeq \psi_{P}\hookrightarrow
\psi_{P}\oplus\psi_{P}\simeq\psi_W \oplus\psi_W =\Psi_D \simeq
\psi_D\oplus\psi_D
\end{equation}
which means that $2^n$ component Dirac spinor is isomorphic to a
$2^n$ component Pauli spinor and that the direct sum of two such
Pauli spinors, equivalent to that of two Weyl spinors may give a
Dirac spinor with $2^{n+1}$ components, which may be then
considered as a doublet of $2^n$    Dirac spinors, which is in
accordance with our programme.

These isomorphisms may be formally represented through unitary
transformations in spinor spaces. In fact indicating with $\G_A$
and $\g_a$ the generators of $\Cl(2n+2)$ and of $\Cl(2n)$
respectively (we adopt the symbols of Appendix A1 corresponding to
$n=2$), assume for the first
$2n$ generators of
$\G_A$ the following forms:
\begin{equation}
\G^{(0)}_a =1_2\otimes \g_a; \ \ \G^{(j)}_a =\s_j\otimes \g_a; \ \
j=1,2,3; \ \ a=1,2\dots 2n
\end{equation}
where $\s_1,\s_2,\s_3$ are Pauli matrices; then if
\begin{equation}
\Psi =\begin{pmatrix} \psi_1\cr \psi_2\end{pmatrix}
\end{equation}
is a $2^{n+1}$-component spinor associated with $\Cl (2n+2)$, the
$2^n$-component spinors $\psi_1$ and $\psi_2$ are:
$$
\begin{array}{rll}
\Cl (2n)-\mbox{Dirac spinors for} &\G^{(0)}_a&\mbox{and we indicate}\
\Psi :=\Psi^{(0)}\\
\Cl_0(2n+2)-\mbox{Weyl spinors for} &\G^{(1)}_a\ \mbox{or}\
\G^{(2)}_a &\mbox{and we indicate}\
\Psi :=\Psi^{(1)}\ \mbox{or}\ \Psi^{(2)}\\
\Cl (2n+1)-\mbox{Pauli spinors for} &\G^{(3)}_a&\mbox{and we
indicate}\ \Psi :=\Psi^{(3)} .
\end{array} 
\eqno(4.6')
$$

Given $\G^{(m)}_a$ with $m=0,1,2,3$, as in eq.(4.5), the other two
generators and the volume element are:
\begin{equation}
\begin{array}{rll}
\G^{(0)}_{2n+1} =\s_1\otimes\g_{2n+1}; &\G^{(0)}_{2n+2}
=\s_2\otimes\g_{2n+1}; &\G^{(0)}_{2n+3} =\s_3\otimes\g_{2n+1}\\
\G^{(1)}_{2n+1} =\s_1\otimes\g_{2n+1}; &\G^{(1)}_{2n+2}
=\s_2\otimes 1; &\G^{(1)}_{2n+3} =\s_3\otimes 1\\
\G^{(2)}_{2n+1} =\s_2\otimes\g_{2n+1}; &\G^{(2)}_{2n+2}
=\s_1\otimes 1; &\G^{(2)}_{2n+3} =\s_3\otimes 1\\
\G^{(3)}_{2n+1} =\s_3\otimes\g_{2n+1}; &\G^{(3)}_{2n+2}
=\s_2\otimes 1; &\G^{(3)}_{2n+3} =\s_1\otimes 1,
\end{array}
\end{equation}
where all the diagonal unit matrices indicated with $1$ have dimension
$2^n$.

Define now the projectors:
\begin{equation}
L:=\frac{1}{2} (1+\g_{2n+1});\quad R:=\frac{1}{2} (1-\g_{2n+1})
\end{equation}
and the unitary transformations:
\begin{equation}
U_j:=1_2\otimes L+\s_j\otimes R=U^{-1}_j;\ \ \ j=1,2,3 .
\end{equation}
Then it is easily seen that
\begin{equation}
U_j\G^{(0)}_AU^{(-1)}_j=\G^{(j)}_A;\ \ \ A=1,2,\dots 2n+2;\ \ \
j=1,2,3
\end{equation}
and therefore
\begin{equation}
U_j\Psi^{(0)}=\Psi^{(j)};\qquad j=1,2,3
\end{equation}
which formally proves the isomorphisms of eq.(4.4).
The above may be easily extended to pseudo-euclidean, in
particular lorentzian, signatures.

Observe that the doublets $\Psi^{(m)}$ are isomorphic according to
eqs.(4.4) and (4.11), however physically inequivalent, since Weyl spinors
(or twistors~\cite{six}) are not invariant for space-time reflections and
therefore
$\psi^W_1,\psi^W_2$ may not be observed as free fermions (if massive), but
may only enter in interactions (weak ones, say). In fact, as we will see
in section 6.3, the $U_j$ of eq.(4.9) will give rise to $SU(2)_L$ of the
electroweak model.

In order to perform the proposed doubling of dimensions of simple
or pure spinor space while increasing also by two the dimension of
the corresponding null-vector space, we need the
following Proposition:\\

\noindent{\bf Proposition 2.}\ Let $\phi_D$ and $\psi_D$ represent
two spinors associated with $\Cl(2n)$, with generators $\gamma_a$,
and let the Weyl spinors $\frac{1}{2}(1\pm\gamma_{2n+1})\psi_D$ be
simple or pure, then, because of Proposition 1, the vectors
$z^\pm\in\C^{2n}$ with components.
\begin{equation}
z^\pm_a=\frac{1}{2}\langle B\phi_D,\gamma_a(1\pm\gamma_{2n+1})
\psi_D\rangle
\end{equation}
are null and their non null sum $z = z^+ + z^-$ with components
\begin{equation}
z_a=z^+_a+z^-_a=\langle B\phi_D,\gamma_a\psi_D\rangle
\end{equation}
is the projection from $\C^{2n+2}$ to $\C^{2n}$ of the vector
$Z\in\C^{2n+2}$ with components
\begin{equation}
Z_a=z_a; \ \ Z_{2n+1}=\langle B\phi_D,\gamma_{2n+1}\psi_D\rangle ;
\ \ Z_{2n+2} =\langle B\phi_D,\psi_D\rangle
\end{equation}
which is null, provided $\psi_D$, conceived as a Weyl spinor of
$\Cl_0 (2n + 2)$, is simple or pure.

The proof is given in ref.\cite{eight}.

The above may be restricted to real spaces of lorentzian
signature:\\

\noindent{\bf Corollary 2.}\   Let $\psi$ represent a spinor
associated with $\Cl(2n - 1, 1)$, with generators $\gamma_a$, and
let its Weyl spinors $\frac{1}{2}(1\pm\gamma_{n+1})\psi$ be simple
or pure, then the vectors $p^\pm\in\R^{2n-1,1}$, with components:
\begin{equation}
p^\pm_a=\frac{1}{2}\tilde\psi\gamma_a(1\pm\gamma_{2n+1})\psi
\end{equation}
where $\tilde\psi =\psi^\dagger\gamma_0$, are null and their non
null sum $p = p^+ + p^-$ is the projection from $\R^{2n+1,1}$ in
$\R^{2n-1,1}$ of a vector $P\in\R^{2n+1,1}$ with real components:
\begin{equation}
P_a=p_a; \ \ P_{2n+1}=\tilde\psi\gamma_{2n+1}\psi ; \ \ P_{2n+2}
=i\tilde\psi\psi
\end{equation}
which is null, provided $\psi$, conceived as a Weyl spinor of
$\Cl_0 (2n + 1, 1)$, is simple or pure.

After this we know that, after the first step of Chapter 3 which gave us
Minkowski signature, our construction will give us only lorentzian
signatures.

Observe that the imaginary unit $i$ in front of $\tilde\psi\psi$
is necessary in order to render $P_{2n+2}$ real for $\psi$ complex
(for $\psi$ real $P_{2n+2}$ is identically null, as may be easily
verified). For the Clifford algebra $\Cl (1,2n-1)$ Corollary 2 is
also true but in eqs.(4.16) we have now to insertÊ
$P_{2n+1}=-i\tilde\psi\gamma_{2n+1}\psi$ and
$P_{2n+2}=\tilde\psi\psi$, both real (see Appendix A1.).

The vector $P\in \R^{2n+1,1}$ with real components $P_A=\{
p_a;P_{2n+1}, P_{2n+2}\}$ will give rise to a Cartan's equation
for the spinor $\Theta$ associated with $\Cl(2n+1,1)$ with
generators $G_A$ (see Appendix A1 from which we adopt the symbols
for $n=3$):
\begin{equation}
P^AG_A(1\pm G_{2n+3})\Theta =0\ ,
\end{equation}
which if
\begin{equation}
\Theta = \begin{pmatrix} \Psi_+\cr \Psi_-\end{pmatrix}
\end{equation}
gives the four equivalent equations:
\begin{equation}
\left( p^a\G^{(m)}_a+P_{2n+1}\G^{(m)}_{2n+1}\pm P_{2n+2}\right)
\Psi^{(m)}_\pm =0;\ \ \  m=1,2,3,0\ ,
\end{equation}
where $\G^{(m)}$ and $\Psi^{(m)}$ are defined by eqs.(4.5) and (4.6$'$)
and
$\Psi^{(m)}$ are isomorphic as shown by eq.(4.11).

We have now the geometrical instruments appropriate to proceed in
our construction consisting in doubling the dimensions of
simple or pure spinor
spaces: $S_+\oplus S_- = S$ and of increasing by two the dimension
of the corresponding null vector spaces: $z_+\oplus z_-\hookrightarrow Z$
or
$p_+\oplus p_-\hookrightarrow P$ at each step.

\section{THE SECOND STEP\\
Majorana equation}

    Let us now sum the two null vectors ${p_\mu}^\pm$ of $\R^{3,1}$
defined by eq.~(3.10):
\begin{equation}
p_\mu = p_\mu^+ + p_\mu^- = \tilde \psi \g_\mu \psi , \qquad
\qquad \mu = 0, 1, 2,3
\end{equation}
because of Corollary 2, they are the projection on $\R^{3,1}$ of a
null vector of a six dimensional space, obtained by adding to the
4 real components $p_\mu$ the following $p_5$ and $p_6$, both
real:
$$
    p_5 = \tilde \psi \g_5 \psi \qquad  p_6 = i\tilde \psi  \psi
        \eqno(5.1')
$$
where, as easily verified, $p_a$  identically satisfy the equation
\begin{equation}
  p_1^2 + p_2^2 + p_3^2 - p_0^2 + p_5^2 + p_6^2 = 0
\end{equation}
that is they build up a null vector in $\R^{5,1}$ and their
directions define the projective Klein quadric equivalent to
$S^4$.

We see that, in accordance with Corollary 2, the signature of the
momentum vector space remains lorentzian. Now since $\Cl(3,1) =\R (4)$ we
may represent it in the space of real spinors $\psi$.  But then $p_6
\equiv 0$ and Cartan's equation becomes:
\begin{equation}
\left(p_\mu \g^\mu + \g_5 m\right) \psi = 0
\end{equation}
that is Majorana equation. For $\psi$
complex instead the equation would be:
\begin{equation}
(p_\mu\gamma^\mu +\gamma_5p_5+ip_6)\psi =0 .
\end{equation}

\section{THIRD STEP: ISOTOPIC SPIN $SU(2)$, THE ELECTRIC
CHARGE AND THE ELECTRO-
WEAK MODEL}

\subsection{Pion-nucleon equation}

Let $\Psi$ represent an 8-component Dirac spinor associated with
$\Cl(5,1)$ generated by $\Gamma_a$, the vectors $p^\pm\in\R^{5,1}$ with
components
\begin{equation}
p^\pm_a=\frac{1}{2}\Psi^\dagger\Gamma_0\Gamma_a(1\pm\Gamma_7)\Psi\
,\qquad a=1,2,\dots , 6
\end{equation}
are null because of Proposition 1 (since $(1\pm\G_7)\Psi$ are simple).
Eqs.(5.1) and (5.1$'$) are a particular case of these (see A1, for
$\Cl(1,5)$ and $\G^{(1)}_a$).

Consider the components: $p_a^+ + p_a^- = p_a =
 \Psi^\dagger \G_0 \G_a \Psi $,
because of Corollary 2, together with the components $p_7 =
\Psi^\dagger \G_0 \G_7 \Psi$ and $p_8 =  \Psi^\dagger \G_0 i\Psi$ they
build up, for $\Psi$ simple spinor of $\Cl_0(7,1)$, a null vector $P \in
\R^{7,1}$ with real components:
\begin{equation}
  p_A = \Psi^\dagger \G_0 \G_A \Psi \quad {\rm where} \quad \G_A =
  \left\{ \G_a, \G_7 ,i\one\right\}\ .  \qquad A=1,\dots,8\,.
\end{equation}
They satisfy (identically) to:
\begin{equation}
p_\mu p^\mu + p^2_5+p^2_6+p^2_7+p^2_8=0
\end{equation}
and the corresponding equations for $\Psi$ may be derived from the
Cartan's equation:
$$p^AG_A(1\pm G_9)\Theta =0    \eqno(4.17')$$
with $G_A$ generators of $\Cl(7.1)$. According to the isomorphisms
discussed in Chapter 4, there are four of them (see eq.(4.19)):
\begin{equation}
\left( p^\mu\G^{(m)}_\mu +p_5\G^{(m)}_5 +p_6\G^{(m)}_6 +p_7\G^{(m)}_7
+M\right)\Psi^{(m)}=0;\ \ \ m=0,1,2,3\ ,
\end{equation}
depending if $\Psi^{(m)}$ is a doublet of Dirac, Weyl, or Pauli
spinors for $m=0; 1,2$; or $3$ respectively. Let us suppose $m=0$ then:
$$
\Psi^{(0)} = \begin{pmatrix} \psi_1 \\ \psi_2 \end{pmatrix}:=N
$$
where $\psi_j$ are Dirac $\Cl(3,1)$-spinors, while $\Gamma_A$ have the
form (see $\G^{(0)}_A$ in eqs.(4.5), (4.7)):
\begin{equation*}
\G^{(0)}_\mu = \begin{pmatrix}\g_\mu & 0 \cr 0 & \g_\mu \end{pmatrix} ,
  \quad \G^{(0)}_5 = \begin{pmatrix}0 & \g_5 \cr \g_5 & 0 \end{pmatrix} ,
  \quad \G^{(0)}_6 = \begin{pmatrix} 0 & -i\g_5 \cr i\g_5 & 0
\end{pmatrix},
 \quad  \G^{(0)}_7 = \begin{pmatrix}\g_5 & 0 \cr 0 & -\g_5
 \end{pmatrix} .
\end{equation*}

We may now interpret eq.(6.4) as an equation of motion by
interpreting $p_\mu$ as generators of Poincar\'e translations and
then substitute $p_\mu$ with $i\frac{\partial}{\partial x_\mu}$,
while defining the spinor $N$ as a function of space-time:
$x_\mu\in M=\R^{3,1}$ (in a similar way as Weyl eq.(3.9$'$) became
(3.11); in this case $\Cl(7,1)$ may be interpreted as a fiber over
$M$), and eq.(6.4) becomes:
\begin{equation}
\left( i\frac{\partial}{\partial x_\mu} \cdot 1 \otimes \g^\mu +
{\vec \pi} \cdot {\vec \s} \otimes \g_5 + M\right) N = 0
\end{equation}
where, according to eq.(6.2):
\begin{equation}
p_\mu =\frac{1}{8}(\tilde\psi_1\gamma_\mu\psi_1+\tilde\psi_2\gamma_\mu\psi_2);\
\ {\vec \pi} = \frac{1}{8}{\tilde N} {\vec \s}\otimes \g_5 N ; \ \
 M = \frac{1}{8} \left({\tilde \psi}_1 \psi_1 + {\tilde \psi}_2
 \psi_2\right)
\end{equation}
with ${\tilde N} = ( \tilde \psi_1, \tilde \psi_2) $. With (6.6)
eq.(6.3) is an identity provided $N$ is simple or
pure~\cite{eight}, and eq.(6.5) is totally spinorial (it may also
become an identity).

Eq.(6.5) is formally identical with the well-known proton-neutron
equation, when interacting with the pseudo-scalar pion isovector,
apart from $M=ip_8$ which here appears to be, like $\vec\pi$,
$x$-dependent (the first option of Chapter 3, that is to conceive $N$
as $p$ dependent is here unacceptable since the Fourier transform of
the equation of motion would give a non local interaction term).

Here it is seen that $p_5,p_6,p_7$ in eq.(6.4) represent
interaction terms with external fields; in this case the pions,
and this will be steadily true, in the following, for all terms
after the first four \footnote{Eight component spinors are subject
to one constraint equation if simple or pure. In this case it
could mean $M=0$ (see eq.(13.1) of ref.\cite{eight}) which, in
turn, because of eq.(6.5) could imply massless pions by which,
notoriously, pion-nucleon physics is greatly simplified: it could
then derive from Cartan's simplicity imposed to the nucleon
doublet.}.

One could also conceive eq.(6.4) as the Fourier transform of an
equation of motion in an 8 dimensional Minkowski space
$M_8=\R^{7,1}$ and define $N$ as a field taking values in $M_8$,
to be after reduced through dimensional reduction; a procedure
here avoided by merely generating ordinary space-time through
$p_\mu$.

As stated our aim is to determine the equations of motion for
fermions conceived as simple or pure spinors where then bosons
appear as external fields bilinearly expressed in terms of
spinors, as in fact appears from eqs.(6.6). At this point it should
be possible to complete the picture by determining the equations of
motion for the pion field, bilinearly expressed in terms of the
nucleon doublet $N$ as in eq.(6.6), in a similar way as Maxwell's
equations may be obtained from the Weyl ones~\cite{seven},\cite{eight}. In
fact E. Fermi and C.N. Yang have shown~\cite{ten} that the pion,
conceived as bilinear (composite) in the nucleon field, is apt to
represent Yukawa theory.

Observe that the pseudoscalar nature of the pion derives from
imposition that $N$ is a doublet of Dirac spinors which, in turn,
imposes the representation $\G^{(0)}_a$ of $\G_a$ where $\g_5$
must be contained in $\G_5$, $\G_6$, $\G_7$ (in order to
anticommute with $\G_\mu$). Again, the fact that the pion field
$\vec{\pi}$ appears here as bilinearly expressed in terms of the
proton-neutron field does not imply that, in the quantized theory,
the pion must be considered as a bound state of proton-neutron.

It is easy to see that if we start from $\Cl(1,3)$ we obtain
$\Cl(1,7)$ and eq.(6.5) assumes the quarternionic form:
$$
(p_\mu\gamma^\mu+s_{4+n}q^n+m)N=0,\quad n=1,2,3 \eqno(6.5')
$$
where $q_n=-i\sigma_n$ are the imaginary quaternion units and
$m=\frac{1}{8} (\tilde\psi_1\psi_1+\tilde\psi_2\psi_2)$ is a real
scalar (see Appendix A1., eq.(A1, 12$'''$)).
The term $s_{4+n}q^n={\vec \pi} \cdot {\vec \s} \otimes \g_5$
presents the so-called isospin internal symmetry  $SU(2)$ of
nuclear forces, here of quaternionic origin.

The group $SU(2)$ is the covering of $SO(3)$ and, traditionally,
isotopic spin symmetry was interpreted as a rotational covariance
in 3-dimensional isotopic spin space. Here instead it is
originated by the generators $\G_5,\G_6,\G_7$ of $\Cl(5,1)$, or
$\Cl(1,5)$ as they appear in the interaction terms:
\begin{equation}
p_5\G_5+p_6\G_6+p_7\G_7 = \vec\pi\cdot\vec\s\otimes\g_5
\end{equation}
of the equation of motion (6.5). Now, as well known, the
generators $\G_a$ of Clifford algebras are reflection operators in
spinor space, from which then internal symmetry should arise.
And this will be true also in the following~\cite{eight}.

W. Heisenberg, the discoverer of isospin symmetry, hoped to derive
it from the conformal group of symmetry, which is notoriously
represented by the group $S0(4,2)$ acting in $\R^{4,2}$, from which
however no compact internal symmetry rotational group may be derived. Instead
if we accept the indication of our derivation, which suggests the
reflection origin of isospin, then, in accordance with
Heisenberg's intuition, there is a correlation with the conformal
group. In fact if we assume $\G_6$ as the second time-like
generator of $\Cl(4,2)$, besides $\G_0$, the generators of the
extra space-time reflections of the conformal group are:
\begin{equation}
\G_5,i\G_6,\G_7
\end{equation}
(where the imaginary unit $i$ in front of $\G_6$ is due to the
fact that the square of a reflection must be equal to the
identity) which are in fact identical (apart from the factor
$\gamma_5$) to the generators appearing in eq.(6.5) or (6.7).
Therefore we may affirm that, according to our derivation, isospin
internal symmetry $SU(2)$ originates from reflections which
identify with those of the conformal group in turn correlated with
the imaginary units of the quaternion field of numbers. This
correlation with conformal reflections might have further
interesting and far reaching implications: see ref.\cite{eight}.

\subsection{Electric charge}

Also the possible origin of electric charge might appear in this third
step. In fact let us write down explicitly eq.~(6.5) in terms of
the two Dirac spinors $\psi_1$ and $\psi_2$ of the $N$ doublet:
\begin{equation}
\begin{array}{rl}
     \left( p_\mu \gamma_\mu + p_7 \gamma_5 + ip_8 \right) \psi_1 +
     \gamma_5 \left( p_5 - i p_6 \right) \psi_2 &= 0\, ,  \\
     \left( p_\mu \gamma_\mu - p_7 \gamma_5 + ip_8 \right) \psi_2 +
     \gamma_5 \left( p_5 + i p_6 \right) \psi_1 &= 0\, .
  \end{array}
\end{equation}
Now all $p_a$ are real, therefore defining
\begin{equation*}
   p_5 \pm i p_6 = \rho e^{\pm i \frac{\omega}{2}}
\end{equation*}
and multiplying the first eq.(6.9) by $e^{i \frac{\omega}{2}}$
we obtain
\begin{equation}
\begin{split}
  \left( p_\mu \gamma_\mu + p_7 \gamma_5 + ip_8 \right) e^{ i
  \frac{\omega}{2}} \psi_1 + \gamma_5 \rho \psi_2 &= 0\,, \\ \left(
  p_\mu \gamma_\mu - p_7 \gamma_5 + ip_8 \right) \psi_2 + \gamma_5 \rho
  e^{i \frac{\omega}{2}} \psi_1 &= 0\,.
\end{split}
\end{equation}
We see then that $\psi_1$ appears with a phase factor $e^{ i
\frac{\omega}{2}}$ corresponding to a rotation through an angle
$\omega$ in the circle defined by
\begin{equation}
  p_5^2 + p_6^2 = \rho^2
\end{equation}
in the vector space of the Klein quadric defined by eq.(6.3),
which in turn corresponds to an imaginary dilation in $\R^{4,2}$.
In fact the corresponding transformation in spinor space is
generated by the Lie algebra element
\begin{equation}
  J_{56} = \frac{1}{2} \left[ \G_5,  \G_6 \right]
\end{equation}
which is obtained from $J_{56}$ in the $SU(2,2)$ covering of the
conformal group, after multiplying the generation $\Gamma_6$ by
the imaginary unit $i$.

Observe that this complexification was intrinsic to the modality
of our construction which brought us to $\Cl(5,1)$ which may be
obtained by setting an imaginary unit factor $i$ in front of the
generator $\G_6$ of $\Cl(4,2)$ as in (6.8) and which finally generated the
$SU(2)$ internal symmetry.

With this interpretation the (non compact) dilation covariance of
the conformal group induced the (compact) $U(1)$
group of symmetry represented by the phase factor in front of
$\psi_1$ and not of $\psi_2$ (or vice-versa).

As stated above we may translate this in the corresponding
Minkowski space-time, and since dilatation covariance is local, we may
consider the phase angle $\omega$ as coordinate dependent $\omega
\rightarrow \omega\left(x\right)$ and then to maintain the
covariance of eq.(6.9) we will have to introduce an abelian gauge
potential $A_\mu$ interacting with $\psi_1$ only and, as easily
seen, the eq.(6.5) will become, in space-time $\R^{3,1}$:
\begin{equation}
  \left\{ \gamma_\mu \left[ i \frac{\partial}{\partial x_\mu} +
  \frac{e}{2} \left( 1 -i \G_5 \G_6 \right) A_\mu \right] + \vec{\pi}
  \cdot \vec{\sigma} \otimes \gamma_5 + M \right\} \begin{pmatrix}
  p\cr n \ \end{pmatrix} = 0
\end{equation}
well representing the equation of the proton-neutron doublet
interacting with the pion and with the electromagnetic potential
$A_\mu$ determining $F_{\mu\nu}$, satisfying eqs.(3.14) in empty space.

Observe that the electric charge $Q$ of the nucleon doublet $N$ is
then represented by:
\begin{equation}
  Q = \frac{e}{2} \left( 1-i \G_5 \G_6 \right),
\end{equation}
where $-i\G_5\G_6=\sigma_3 \otimes 1$, that is the third component
of isospin generator. Furthermore the proton $p$ and the neutron
$n$ are eigenstates of $-i\G_5\G_6$ corresponding to the
eigenvalues $+1$ and $-1$, respectively (see also ref.\cite{eight}).

Quaternions might have a role also in the origin of the
electroweak model where also a doublet of (left-handed) fermions
appears, as we will see in the next section.

\subsection{The electroweak model}

Let us go back to eq.(6.4) representing four equivalent equations of
motion for the fermion doublet $\Psi^{(m)}$ labelled by the index
$m=0,1,2,3$. In section 6.1 we adopted $m=0$ corresponding to a doublet
of Dirac $\Cl(3,1)$ spinors, obtaining $SU(2)$ symmetric
pion-nuclear equation (6.5). For $m=1,2,3$,
corresponding to Weyl and Pauli doublets, we may expect to obtain another
$SU(2)$ symmetric equation: it will be the one of the electroweak
model.

In fact the
corresponding spinors  $\Psi^{(m)}$
are isomorphic, however, for what is observed in Chapter 4, they are
not physically equivalent.
Suppose now that the doublet $\Psi^{(m)}$ represents a
lepton pair whose electroweak charge derives from a local $U(1)$ phase
covariance of
$\Psi^{(m)}$, (as we will see in Chapter 7). Then a minimal coupling
(or covariant derivative in space time) will impose a gauge vector
field $A^{(m)}_\mu$ and (6.4) will become:
\begin{equation}
\left[(p^\mu-A^\mu_{(m)})\Gamma_\mu^{(m)}+\sum^7_{j=5} p_j
\Gamma^{(m)}_j +M\right]\Psi^{(m)}=0,\quad m=0,1,2,3
\end{equation}
where:
\begin{equation}
A^{(m)}_\mu = \widetilde{\Psi}^{(m)}\Gamma^{(m)}_\mu\Psi^{(m)} .
\end{equation}

Let us now consider the unitary operators $U_j$ defined in
eqs.(4.9) where now $L$ and $R$ are the chiral projectors:
\begin{equation}
L=\frac{1}{2}(1+\g_5);\quad R=\frac{1}{2}(1-\g_5) .
\end{equation}

The operators $U_j$ represent the isomorphisms of eqs.(4.4)
because of which:
\begin{equation}
U_j\Psi^{(j)}=\Psi^{(0)};\quad j=1,2,3\ .
\end{equation}

Let us now indicate with $\Psi^{(m)}_L$ the left-handed part of
$\Psi^{(m)}$:
\begin{equation}
\Psi^{(m)}_L=1_2\otimes L\Psi^{(m)}=\begin{pmatrix}
\psi_{1L}\cr \psi_{2L}\end{pmatrix}
\end{equation}
then, since the $U_j$, given in (4.9), have in common the term
$1_2\otimes L$, as a consequence of (6.18), we have that
\begin{equation}
\Psi^{(m)}_L=\Psi^{(m')}_L
\end{equation}
for all four values of $m$ and $m'$.

Now since $\Psi^{(m)} =\Psi^{(m)}_L+\Psi^{(m)}_R$ we have that,
summing eqs.(6.15) for the indices $m=1,2,3$ they will give rise,
reminding that $\G^{(j)}_\mu =\s_j\otimes\g_\mu$ (see eqs.(4.5)),
to the equation:
\begin{equation}
\sum^3_{k=1}\ \left( p^\mu\G^{(k)}_\mu +\sum^7_{j=1}\
p_j\G^{(k)}_j+M\right) \Psi^{(k)} -\sum^3_{k=1}\
A^\mu_{(k)}\G^{(k)}_\mu\Psi^{(k)}_R-
\vec{A}^\mu\cdot\vec\sigma\otimes\gamma_\mu\Psi^{(0)}_L=0\ ,
\end{equation}
where the last term presents an $SU(2)_L$ internal gauge symmetry (this
time in the dynamical sector of the equation).

If we suppose that the lepton doublet $\Psi^{(0)}$ represents an
electron $e$ and a massless, left-handed neutrino $\nu_L$:
$$
\Psi^{(0)} =\begin{pmatrix} e\cr \nu_L\end{pmatrix} ,
$$
we will have that, because of eq.(6.20):
\begin{equation}
\Psi^{(m)}_L =\Psi^{(0)}_L =\begin{pmatrix} e_L\cr
\nu_L\end{pmatrix};\ \ \ \Psi^{(m)}_R =e_R, \ \ \ m=0,1,2,3
\end{equation}
and, after summing eqs.(6.15),
taking into account that, for leptons, not possessing strong charges, the
terms
$p_5,p_6,p_7$ will be absent, we easily obtain:
\begin{equation}
(p_\mu\g^\mu +M)\begin{pmatrix} e\cr \nu_L\end{pmatrix} -\vec
A_\mu\cdot\vec\s\times\g^\mu\begin{pmatrix} e_L\cr
\nu_L\end{pmatrix} +(B_\mu\g^\mu+\tau )e_R=0
\end{equation}
where:
$$\vec A_\mu=\widetilde{\Psi}^{(0)}\vec\s\otimes\g_\mu\Psi^{(0)} ,
$$
$B_\mu$ is an isosinglet four-vector and $\tau$ a scalar.
Eq.(6.23) represents the geometrical starting point for the
electroweak model.

\subsection{Dirac's equation}

Both from eq.(6.5) and from eq.(6.23) we may obtain the free
Dirac's equation; in fact we need only to eliminate from (6.5) the pion
interaction terms and what remains is the Dirac's equation for the
proton and the neutron; similarly from (6.23) eliminating the
electroweak interactions we obtain the Dirac's
equation for the free electron. We see then that the free Dirac equation
is only conceivable as an approximate equation when interactions may
be ignored, which is natural since the Dirac spinor is complex.
The only exact equation of motion, for four component spinors which naturally
emerges in our construction is the Majorana one: eq.(5.3).

\section{FOURTH STEP: THE STRONG (OR ELECTROWEAK) CHARGE}

\subsection{The baryon-lepton quadruplet}

The construction may be continued observing that the real null
vector with components $p_A$ given by eq.~(6.2) for $N$, thought
as a Weyl spinor of $\Cl_0(7,1)$, may be considered as a
particular case of the following:
\begin{equation}
  P_A^\pm = \Theta^\dagger G_0 G_A \left( 1 \pm G_9 \right)
  \Theta\,, \qquad A=1,2,\dots,8 \,,
\end{equation}
where $\Theta$ is a sixteen component spinor associated with
$\Cl(7,1)$ of which $G_A$ are the generators and $G_9$  the volume
element.

Again $\Theta$ may be considered in the Dirac basis
\begin{equation}
  \Theta = \begin{pmatrix} N_1 \cr N_2 \end{pmatrix}
\end{equation}
where $N_1$ and $N_2$ are Dirac spinors of $\Cl(5,1)$ and, again,
defining
\begin{equation}
  P_A = P_A^+ + P_A^- = \Theta^\dagger G_0 G_A \Theta \qquad
  A=1,2,\dots,8
\end{equation}
and
\begin{equation}
\begin{split}
  P_9 &= \Theta^\dagger G_0 G_9 \Theta \\
  P_{10} &= i\Theta^\dagger G_0 \Theta
\end{split}
\end{equation}
we obtain, for $\Theta$ simple or pure, the components $P_\alpha$
of a ten dimensional real null vector which defines the Cartan's
equation for the spinor $\Phi$, associated with $\Cl(9,1)$, with
generators $\cg_\alpha$:
$$
P_\alpha\cg_\alpha (1\pm \cg_{11})\Phi =0\qquad \alpha =1,2,\dots
, 10     \eqno(4.17'')
$$
from which the equation for the 16-component spinor $\Theta$ may
be derived:
\begin{equation}
  \left( P_aG^a + P_7 G_7 + P_8 G_8
  + P_9 G_9 + i P_{10} \right) \Theta =0 \,,
\end{equation}
where $a=1,2\dots 6$. For $\Theta$ in the Dirac basis we may, with
the same procedure as that of section 6.2, show that the Dirac
spinors $N_1$ and $N_2$ obey a system of equations where $N_1$ (or
$N_2$) is multiplied by a phase factor $e^{i \tau/2}$, where the
angle $\tau$ represents a rotation in the circle
\begin{equation}
   P_7^2 + P_8^2 = \rho^2
\end{equation}
for which eq.~(7.5) is covariant (in spinor space it is generated
by $G_7G_8$). This $U(1)$ symmetry of $N_1$ (or $N_2$) may be
interpreted as a charge which, being different from the electric
charge (generated by $\G_5\G_6$), could be the charge of strong
forces. In this case then, $N_1 = \begin{pmatrix} \psi_1 \cr
\psi_2
\end{pmatrix} $ could represent the nucleon doublet while $N_2 =
\begin{pmatrix} \psi_3 \cr \psi_4 \end{pmatrix} $ the electron
neutrino doublet say, both of which contain an electrically
charged and neutral component. Then it is easy to see that, since
$N_1$ and $N_2$ are eigenstates of $-iG_7G_8$ corresponding to the
eigenvalues $+1$ and $-1$ respectively, while the Dirac spinor
components of $N_1$ and $N_2$ are eigenstates of $-i\G_5\G_6$,
again corresponding to the eigenvalues $+1$, $-1$, if the charged
partner of the nucleon doublet $N_1$ (the proton) has the charge
$+e$, the charged partner of the lepton doublet $N_2$ (the
electron) should have the charge $-e$ (see ref.\cite{eight}), as it
happens, in fact.

As seen above in each of the doublets quaternions generate an internal
symmetry which could then be $SU(2)$ isospin for $N_1$ and $SU(2)_L$
electroweak for $N_2$. Suppose now that, as discussed in section 6.2, the
rotation in the circle (7.6) generating $U(1)$ is local, then it will
generate a gauge interaction, like it did in eq.(6.13); however now,
because of those $SU(2)$ it will be a non abelian gauge field theory (and
this time, despite the reflection origin in spinor space, those $SU(2)$
may be conceived as covering of rotation groups, as will be discussed
elsewhere), obviously to be completed, with the dynamical part of
Yang-Mills.

\section{THE FIFTH AND LAST STEP}

 Continuing the construction, the $P_\alpha$ of eqs.(7.3) and
(7.4) are a particular realization of the following:
\begin{equation}
 P^\pm_\alpha = \Phi^\dagger \cg_0 \cg_\alpha \left( 1 \pm \cg_{11} \right)
 \Phi \qquad \alpha=1,2,\dots,10 .
\end{equation}
Then, for $\Phi$ simple (as Weyl of $\Cl_0(11,1)$)
\begin{equation}
 P_\alpha = P_\alpha^+ + P_\alpha^- = \Phi^\dagger \cg_0 \cg_{\alpha} \Phi
\end{equation}
together with
\begin{equation}
  P_{11} = \Phi^\dagger \cg_0 \cg_{11} \Phi \qquad  P_{12} =
  i\Phi^\dagger \cg_0 \Phi
\end{equation}
build up the components of a 12 dimensional real null vector
defining the Cartan's equation
\begin{equation}
  \left( P_A \cg^A + P_9 \cg_9 +  P_{10} \cg_{10} +  P_{11}
  \cg_{11} + i P_{12} \right) \Phi =0 ,\ \ \ A=1,2,\dots,8 .
\end{equation}

\noindent For $\Phi$, in the Dirac spinor representation
\begin{equation}
  \Phi = \begin{pmatrix}  \Theta_1 \cr \Theta_2 \end{pmatrix}
\end{equation}
where $\Theta_1$ and $\Theta_2$ may be considered  as Dirac
spinors of $\Cl(7,1)$, eq.(8.4) will present, with the same
procedure as that of section 6.2 and Chapter 7, an $U(1)$ phase invariance
of
$\Theta_1$ generated by $\cg_9 \cg_{10}$, corresponding to a
rotation through an angle $\s$ in the circle
$P_{9}^2+P_{10}^2=\rho^2$. We may interpret it as the $U(1)$
corresponding to a strong charge (or hyper\-charge). Then the
quadruplet of fermions contained in $\Theta_1$ and $\Theta_2$
could represent baryons and leptons respectively, and each one of
them obeys in principle to an equation like (7.5).

Clifford algebras $\Cl(k,l)$ obey the notorious periodicity
theorem:
\begin{equation}
   \Cl( l+4,k+4 ) = \Cl(l+8,k) = \Cl(l,k+8) = \Cl(l,k) \otimes
   {\mathcal R}(16)
\end{equation}
where ${\mathcal R}\left(16\right)$ stands for the algebra of $16
\times 16$ real matrices. Therefore, the maximal Clifford algebra
to study in our construction will be the algebra
\begin{equation}
  \Cl(9,1) = {\mathcal R}(32) =   \Cl(1,9)\,,
\end{equation}
admitting real Majorana-Weyl spinors, since after this the
geometrical structures, because of the periodicity theorem, will
repeat themselves, and, furthermore the number of constraint
equations becomes too high: $66$ and $364$ for $\Cl (11,1)$ and
$\Cl(13,1)$, respectively.

Before examining the possibility of representing with $\Theta_1$
and $\Theta_2$ baryons and leptons respectively, we wish first to
define dimensional
 reduction procedure in our formulation and to study its possible physical
meaning.

\section{DIMENSIONAL REDUCTION}

In our approach, dimensional reduction will simply
consist in reversing the steps of our construction. Precisely if
$\Phi$ represents a $2^n$ component Dirac spinor associated to the
Clifford algebra $\Cl (2n-1,1)$, with generators
$\g_a:\{\g_1,\g_2,\dots\g_{2n}\}$ we may reduce the spinor $\Phi$
to the $2^{n-1}$ component Weyl spinors $\varphi^W_\pm$ through
the projectors $\pi^{(1,2)}_\pm = \frac{1}{2} (1\pm\g_{2n+1})$
see~\cite{eight}:
\begin{equation}
\pi ^{(1,2)}_\pm:\ \ \Phi\to\frac{1}{2} (1\pm\g_{2n+1})\Phi =\varphi^W_\pm
.
\end{equation}
Correspondingly the $2n+2$ dimensional null vector with
components:
\begin{equation}
P_A=\Phi^\dagger\g_0\g_A\Phi\qquad A=1,2,\dots 2n+2
\end{equation}
where $\g_A:\{\g_a,\g_{2n+1},i\one\}$ will be
reduced to:
\begin{equation}
\pi ^{(1,2)}_\pm:\ \ P_A\to
p^\pm_a=\Phi^\dagger\g_0\g_a(1\pm\g_{2n+1})\Phi ,\quad a=1,2,\dots 2n
\end{equation}
since, according to Corollary 2, $P_{2n+1}\equiv 0\equiv
P_{2n+2}$.

However there are also other possibilities. In fact according to the
isomorphisms indicated in (4.4) we may also reduce the Dirac spinor $\Phi$
to $2^{n-1}$ components Dirac spinors
$\varphi^D_\pm$  associated with the Clifford algebra $\Cl
(2n-3,1)$. It is easy to see~\cite{eight} that in this case we have to use
the projectors $\pi^{(0)}_\pm=\frac{1}{2} (1\pm
i\g_{2n-1}\g_{2n})$:
\begin{equation}
\pi^{(0)}_\pm:\ \ \Phi\to\frac{1}{2} (1\pm i\g_{2n-1}\g_{2n})\Phi
=\varphi^D_\pm
\end{equation}
and correspondingly we have $P_{2n-1}\equiv 0\equiv P_{2n}$; that
is
\begin{equation}
\pi^{(0)}_\pm:\ \ P_A\to p^\pm_b = \Phi^\dagger\g_0\g_b(1\pm
i\g_{2n-1}\g_{2n})\Phi
\end{equation}
where $b=1,2,\dots 2n-2,2n+1$.

According to the isomorphisms indicated in (4.4) we may also
reduce the spinor $\Phi$ to Pauli spinors associated with $\Cl
(2n-2,1)$. It is easy to see~\cite{eight} that in this case we have to use
the projectors: $\pi^{(3)}_\pm =\frac{1}{2}(1\pm
i\g_{2n}\g_{2n+1})$:
\begin{equation}
\pi^{(3)}_\pm:\ \ \Phi\to\frac{1}{2}(1\pm i\g_{2n}\g_{2n+1})\Phi
=\varphi^P_\pm
\end{equation}
as a consequence of which: $P_{2n}\equiv 0\equiv P_{2n+1}$.

We see then that there are more possibilities of dimensional
reduction each of which halves the dimension of spinor space and
reduces by two the dimension of momentum space. The reduced spinor
spaces are isomorphic because of the isomorphisms represented
in (4.4), (4.11) while, what changes in each
reduction, are the two components of momentum space, appearing in
the interaction terms, which are eliminated from the equations of
motion (see also ref.\cite{eight}), therefore the reduced spinors may
result physically not equivalent, and this inequivalence in fact may give
rise to the existence of families as we will see in the next Chapter.

Dimensional riduction will eliminate $P_A$ with $A\geq 5$ and then
eliminate the interaction terms from the equation of motion, which in
turn will mean descending from higher to lower energy phenomena.
$P_\mu$ with
$\mu =1,2,3,0$ will represent the dynamical term, which when interpreted
as Poincar\'e translation, generate space-time, which in this approach is,
and remains, four dimensional.

\section{BARYONS AND LEPTONS}
\subsection{Baryons}

Let us now assume in eq.(8.5) the 16 component spinor
 $\Theta_1$ to be of the form:
\begin{equation}
 \Theta_1 = \begin{pmatrix} b_1 \cr b_2 \cr b_3 \cr b_4 \end{pmatrix}: =
 \Theta_B \label{11.1}
\end{equation}
where $b_j$ are $\Cl(3,1)$-Dirac spinors, representing a quadruplets of
fermions or quarks, presenting strong charge represented by its
$U(1)$ covariance generated by $\cg_9\cg_{10}$. Suppose it obeys eq.(7.5)
with $P_\alpha$ given by (7.3), (7.4), being the components of a
null vector satisfy identically to:
\begin{equation}
 -P_\mu P^\mu = P_5^2+P_6^2+P_7^2+P_8^2+P_9^2+P_{10}^2=\cm^2 ,
 \label{11.2}
\end{equation}
where $\cm$ is an invariant mass. The directions of $P_5,P_6,\dots
P_{10}$ define $S^5$. Therefore it is to be expected that the
quadruplets may present a maximal $SU(4)$ internal symmetry
(covering group of $SO(6)$), which will be the obvious candidate
for the maximal possible flavor internal symmetry. The most
straightforward way to set it in evidence is to determine the 15
generators if the Lie algebra of $SU(4)$ (determined by the
commutators $\left[G_i,G_k\right]$ for $5 \leqslant j,k \leqslant
10$) represented by $4 \times 4$ matrices, whose elements are
either $1_4$ or $\g_5$, acting in the space of the $\Theta_B$
quadruplet. Let them be $\lambda_j$; where $1\leqslant j \leqslant
15$. Denote with $f_j$, $1\leqslant j \leqslant 15$ the components
of an emisymmetric tensor building up an automorphism space of
$SO(6)$. Then, a natural\footnote{In a similar way as $\left(
i\partial_\mu \g^\mu + \left[ \g_\mu , \g_\nu\right]F^{\mu\nu} +m
\right) \psi =0$, where $F_{\mu\nu}$ is the electromagnetic
tensor, is a natural equation for a fermion, the neutron say,
presenting an anomalous magnetic moment.} equation of motion for
$\Theta_B$ could have the general form:
\begin{equation}
  \left( P_\mu G^\mu + \sum^3_{j=1}\lambda_jf^j+\sum_{j=4}^8
\lambda_j f^j + \sum_{j=9}^{15}
  \lambda_j f^j \right)\Theta_B =0
\end{equation}
 where $\lambda_1\lambda_2\lambda_3$ are the generators of
$SU(2)$; $\lambda_1,\dots ,\lambda_8$ those of $SU(3)$ and
$\lambda_1,\dots ,\lambda_{15}$ those of $SU(4)$.

 In order to study the possible physical information contained in the
 quadruplet $\Theta_B$, let us now operate with our dimensional
 reduction.

 The most obvious will be to adopt the projector $\pi^{(1,2)}_\pm =
 \frac{1}{2} \left(1
 \pm G_9 \right)$ which is the image in $\Cl(9,1)$ of $\frac{1}{2}
 \left(1 \pm \cg_9 \cg_{10}\right)$ of $\Cl(1,11)$, and, as we
 have seen in
 Chapter 8, $\cg_9\cg_{10}$ was the generator of $U(1)$ corresponding
 to the strong charge. Now, the reduction
\begin{equation}
  \Theta_B \to \frac{1}{2} \left( 1 \pm G_9 \right) \Theta_B = N_\pm
\end{equation}
implies that
\begin{equation}
  p_{9,10}^\pm = \Theta_B^\dagger G_0  G_{9,10} \left(1 \pm G_9\right)
  \Theta_B \equiv 0  \,.
\end{equation}
Therefore the null vector $P_\alpha$ given by (7.3) and (7.4)
reduces to $p_A$ given by eq.(6.2), which means that $N_+$ or
$N_-$, conceived as a doublet of fermions obeys eq.(6.5) of the
nucleon doublet say, presenting an $SU(2)$ isospin internal symmetry
generated by quaternions. In this way we may realize the first
term, after the dynamical one $P_\mu G^\mu$, in eq.(10.3). But we
know that at least also the second one representing $SU(3)$
symmetry should be possible. In the next Chapter we will see that
it may emerge from octonions.

\subsection{Leptons and families}

We will now interpret the 16-component spinors $\Theta_2$ in
eq.(8.5) as a quadruplet of leptons:
$$
\Theta_2:=\begin{pmatrix} \ell_1\cr \ell_2\cr \ell_3\cr \ell_4
\end{pmatrix} :=\Theta_{\cal L} .
\eqno(10.1')
$$

We may now operate the dimensional reduction of $\Theta_{\cal L}$
with the projector $\pi^{(1,2)}_\pm =\frac{1}{2} (1\pm G_9)$ as we did
for $\Theta_B$:
$$\Theta_{\cal L}\to\frac{1}{2} (1\pm G_9)\Theta_{\cal L}=\varphi^W_\pm
\eqno(10.4')
$$
and consequently $p^\pm_9\equiv 0\equiv p^\pm_{10}$. Now, at
difference with the baryon quadruplet, for leptons, dimensional
reduction, equivalent to decoupling, is necessary, since the
leptons, not presenting strong change, may neither obey eq.(7.5)
nor to the full eq.(10.3); and then the doublet $\varphi^W_+$ say,
may only present the $SU(2)_L$ symmetry of the electroweak model,
as shown in section 6.3.

The eight component spinors $\varphi^W_\pm$ are Weyl spinors and
therefore the first four generators $G_\mu$ of $\Cl (7,1)$, which
determines their physical behaviour, or transformation properties
with respect to the Poincar\'e group will be:
\begin{equation}
G^{(1,2)}_\mu =\sigma_{1,2}\otimes\G_\mu
\end{equation}
where $\G_\mu$ are the first four generators of $\Cl (5,1)$. These
in turn, which determine the physical behaviour of the fermions in the
doublet, will have, according to eqs.(4.5), the following possible forms:
$$
\G^{(1,2)}_\mu =\sigma_{1,2}\otimes\g_\mu,\ \ \G^{(3)}_\mu
=\sigma_3\otimes\g^\mu\ \ {\rm or }\ \ \G^{(0)}_\mu =\one\otimes\g_\mu
\eqno(10.6')
$$
depending if we have a doublet of Weyl, Pauli or Dirac
spinors respectively.

We may also conceive $\Theta_{\cal L}$ as a doublet of Pauli spinors,
and consequently reduce it with the projector $\pi^{(3)}_\pm =\frac{1}{2}(1\pm
iG_8G_9)$:
$$
\Theta_{\cal L}\to\frac{1}{2}(1\pm iG_8G_9)\Theta_{\cal L}=\varphi^P_\pm
\eqno(10.4'')
$$
because of which $p^\pm_8\equiv 0\equiv p^\pm_9$ and the generators $G_\mu$
will be:
\begin{equation}
G^{(3)}_\mu =\sigma_3\otimes\G_\mu
\end{equation}
and, again  $\G_\mu$ may have the forms (10.6$'$).

For $\Theta_{\cal L}$  doublet of Dirac spinors we will adopt the
projector $\pi^{(0)}_\pm =\frac{1}{2}(1\pm iG_7G_8)$:
$$
\Theta_{\cal L}\to\frac{1}{2}(1\pm iG_7G_8)\Theta_{\cal
L}=\varphi^D_\pm
\eqno(10.4''')
$$
and then: $p^\pm_7\equiv 0\equiv p^\pm_8$
while:
\begin{equation}
G_\mu^{(0)} =\one\otimes\G_\mu
\end{equation}
and again $\G_\mu$ may have the forms (10.6$'$).

We see then through the projectors $\pi^{(1,2)}_\pm ,\pi^{(3)}_\pm
,\pi^{(0)}_\pm$, there is the possibility of reduction of
$\Theta_{\cal L}$ to an 8-component Weyl, Pauli or Dirac spinor
respectively (correlated with quaternions) which are isomorphic
according to (4.4) and (4.11). Not equivalent
are instead the corresponding dimensional reduction in momentum
space since the eliminated components of momentum will be:
$p_{10},p_9;p_9p_8$ and $p_8,p_7$ for the Weyl, Pauli and Dirac
case respectively. Now, since these components appear in the
equations of motion as interaction terms, the three reduced lepton
doublets will obey different equations of motion obtained from eq.(7.5)
eliminating the corresponding pair of terms. Correspondingly the invariant
mass equation (10.2) will reduce to:
\begin{eqnarray}
-p_\mu p^\mu &=& p^2_5+p^2_6+p^2_7+p^2_8 = m^2_W \nonumber \\
-p_\mu p^\mu &=& p^2_5+p^2_6+p^2_7+p^2_{10} = m^2_P\\
-p_\mu p^\mu &=& p^2_5+p^2_6+p^2_9+p^2_{10} = m^2_D \nonumber
\end{eqnarray}
for the Weyl, Pauli and Dirac case respectively.

Dimensional reduction of the lepton quadruplet brings us then to 3
fermion doublets of charged-neutral leptons with different
invariant masses $m_W,m_P,m_D$, each one of them lower than the
one $M$ of eq.(10.2) of the baryon multiplet, reminding the 3
families of leptons discovered in nature.

In our imbedding of spinors and compact momentum spaces in higher
dimensional ones we could represent physical phenomena whose
characteristic energy and momentum steadily increased with the
dimension of the spaces. One could then expect that the mean
values of the components with higher indices (9,10) should be
higher than those with lower ones (5,6). In this expectation then
the $e,\mu,\tau$ leptons could then be identified with
$\varphi^W,\varphi^P,\varphi^D$ respectively.

We could obviously repeat the same reductions with the baryon
quadruplet $\Theta_B$  and we would obtain 3 families of quarks to
be correlated with the corresponding 3 families of leptons, and
with similar properties, as it effectively appears in natural
phenomena.

The lepton quadruplet $\Theta_{{\cal L}}$ presents a further
problem. In fact we know that because of the absence of strong
charge it has to be reduced to the doublets $\varphi_\pm$  given in
(10.4$'$), (10.4$''$) and (10.4$'''$) and we also know that $G_7G_8$
generates an
$U(1)$ phase invariance for $\varphi_+$ say, which could be interpreted
as the electroweak charge. But then the fermion doublets represented by
$\varphi_-$ should be free from it. Their fermions could then obey to
eq.(5.3) (or eq.(5.4)) that is they could be Majorana spinors, only
subject to gravitational forces. Furthermore if we trust the above
picture, they should be abundant in nature as leptons are. What
could then come naturally to mind is dark matter.

Baryons present notoriously, besides the internal symmetry $SU(2)$
possibly of quaternionic origin, also the internal symmetry
$SU(3)$ both of flavour and of color. We will show that it might
originate from the third and last division algebra; that of
octonions.

\section{OCTONIONS}

\subsection{In Clifford algebras}

The Clifford algebra $\Cl (9,1)$ or $\Cl (1,9)$ may be represented
with octonions; in fact it is known~\cite{eleven} that:
\begin{equation}
\Spin (9,1) = \Spin (1,9) = SL(2,{\bf o})= SL(32,\R )
\end{equation} where ${\bf
o}$ stands for octonions (see Appendix A2).  Independently, but
coherently with our approach, a spinor equation of motion in a ten
dimensional momentum space $\R^{1,9}$ was proposed by T. Dray and
C.A. Manogue~\cite{twelve} of the form:
\begin{equation}
P\Theta := \begin{pmatrix} p_{10}+p_9 & p_8 -
\mathop{\sum}\limits_{j=1}^7 p_j e_j \cr p_8 +
\mathop{\sum}\limits_{j=1}^7 p_j e_j & p_{10}-p_9 \end{pmatrix}
\begin{pmatrix}
{\bf o}_1 \cr {\bf o}_2
\end{pmatrix} = 0
\end{equation}
where $e_j$ are the octonion units and ${\bf o}_1,{\bf o}_2$
octonions.

Let us now write Cartan's eq.(7.5), for the baryon and lepton
quadruplets $\Theta_B$ and $\Theta_{\cal L}$  discussed in the
previous chapter, in matrix form. Adopt for the generators $G_A$
of $\Cl (7,1)$ the representation:
$$
G_a=\s_2\otimes\G_a; \ \ G_7=\s_2\otimes\G_7, \ G_8=\s_1\otimes 1,
\ G_9=\s_3\otimes 1,\ \ \ a=1,2,\dots 6
$$
and we obtain:
$$
P \Theta_{\stackrel{\scriptstyle{B}} {\scriptstyle{\cal L}}  } =
\begin{pmatrix} \pm ip_{10} + p_9 & p_8 -i
\mathop{\sum}\limits_{j=1}^7 p_j
  \G_j \cr p_8 + i
\mathop{\sum}\limits_{j=1}^7 p_j \G_j & \pm i p_{10} -p_9 \end{pmatrix}
  \begin{pmatrix} \varphi^W_+ \cr \varphi^W_- \end{pmatrix} =0 .
\eqno(11.2')
$$

Comparing it with (11.2) we see that $\varphi^W_+ , \varphi^W_-$
substitute the octonions ${\bf o}_1,{\bf o}_2$ and the seven generators
$i\G_j$ the octonion units $e_j$. The imaginary unit $i$ in front
of $p_{10}$ simply means that while in (11.2) $p_{10}$ is the
energy or time-like direction, in (11.2$'$) it is $p_{0}$. The $+$
and $-$ sign in front of $p_{10}$ refer to the Baryon and Lepton
multiplets respectively (equivalent forms of eq.(11.2$'$) are
given in Appendix A1 for the signature $1,9$).

Eqs.(11.2) and (11.2$'$) should be equivalent because of
eqs.(11.1), and in fact also in ref.\cite{twelve}, through reduction from
$10-$ to $4-$dimensional momentum space, the equations of motion
for three families of lepton -- massless neutrino pairs are
obtained as correlated with the quaternion units.

The privilege of eqs.(11.2$'$) is that their physical meaning is
transparent (one could easily transform them in Minkowski
space-time and then build up there the corresponding familiar
Lagrangian formalism). With eq.(11.2) instead one may hope to
discover the possible role in physics of octonions after the one
of quaternions. In fact the group
of automorphism $G_2$ of octonions has a subgroup $SU(3)$, once a
preferred direction, or octonion unit, is selected. However with
eq.(11.2) one has to pay the price of non associativity of the
algebra of octonions. To overcome this difficulty a matrix
representation of octonions could be helpful in order to deal with
them in Clifford algebras. We will try to find one in the frame of
$\Cl (2,3)$; the anti De Sitter Clifford algebra.

Let us in fact assume
\begin{equation}
\g_n=\begin{pmatrix} 0 &-\s_n\cr \s_n &0\end{pmatrix};\ \
\g_0=\begin{pmatrix} 0 &1\cr 1 &0\end{pmatrix}\ \
\g_5=\begin{pmatrix} 1 &0\cr 0 &-1\end{pmatrix}
\end{equation}
where $n=1,2,3$, as generators of $\Cl (2,3)$. Then, if
$e_1,e_2,\dots e_7$ are the seven imaginary units of octonions
and $e_0$ the unit of the algebra, we define, following J. Daboul
and R. Delbourgo~\cite{thirteen}:
\begin{equation}
e_0:=1_4,\ e_n:=\g_n;\ \ e_7:=i\g_5 ;\ \ \hat
e_n:=e_{n+3}:=e_ne_7=i\g_n\g_5 .
\end{equation}
In ref.\cite{thirteen} it is shown that (11.4) close the octonion algebra
after an appropriate modification of the rule of matrix
multiplication. We will instead indicate with the symbol $\odot$
the product of the first six octonion units defined by
\begin{equation}
e_j\odot
e_k:=\delta_{jk}e_je_k+(1-\delta_{jk})i\g_5\g_0e_j\g_0e_k\g_0
\end{equation}
where $j,k=1,2,\dots 6$, and, on the r.h.s. all products mean
standard matrix multiplication, then (11.5) taking into account of
(11.4) which defines the products with the seventh, $e_7$, close the octonion
algebra (see A2).

\subsection{$SU(3)$ flavor}

Let us now consider $\Cl (1,7)$ and the corresponding Cartan's
equation:
\begin{equation}
P_AG^A(1\pm G_9)\Theta =0\qquad A=1,2,\dots 8
\end{equation}
and adopt for $G_A$ the representation (see (A1.10)):
\begin{equation}
G^{(2)}_A:\{ G_a=\s_2\otimes\G_a;\ \ G_7=-i\s_2\otimes\G_7, \
G_8=-i\s_1\otimes 1_8,\ G_9=\s_3\otimes 1_8\}\ ,
\end{equation}
where $a=1,2,\dots 6$;
then, if $\G_a$ and $\G_7$ are assumed in the Dirac representation
$\G^{(0)}_a$ and $\G^{(0)}_7$ (see A1.7) we have:
\begin{equation}
G_{5,6,7} =\begin{pmatrix} 0&-\s_{1,2,3}\cr \s_{1,2,3}
&0\end{pmatrix} \otimes\g_5
=-i\sigma_2\otimes\sigma_{1,2,3}\otimes\g_5 .
\end{equation}
Therefore we may define:
\begin{equation}
G_{4+n}:=e_n\otimes\g_5; \ \ iG_9:=e_7\otimes 1_4; \ \
iG_{4+n}G_9:=\hat e_n\otimes\g_5 .
\end{equation}

In this way we get a definition of the octonion unit elements in
$\Cl (1,7)$ which with the $\odot$ product close the octonion
algebra (see A2).

Similarly for $\Cl (1,9)$ with generators $\cg_\alpha$  and
Cartan's equation
\begin{equation}
P_\alpha \cg^\alpha (1\pm \cg_{11})\Phi =0\qquad \alpha = 1,2,\dots 10
\end{equation}
we find that (see $\cg^{(2)}_A$ in A1.15):
\begin{equation}
\cg_{7,8,9} =-i\s_2\otimes\s_{1,2,3}\otimes\G_7
\end{equation}
and then the octonion units may be defined by:
\begin{equation}
\cg_{6+n}:=e_n\otimes\G_7;\ \ i\cg_{11}:=e_7\otimes 1_8;\ \
i\cg_{6+n}\cg_{11}:=\hat e_n\otimes\G_7
\end{equation}
which, with the above conventions, close the octonion algebra.

Observe that there may be other representations of the octonion
units in the generators of $\Cl (1,7)$ and $\Cl (1,9)$ (see A2).

Let us now define the so-called complex octonions:
\begin{equation}
u_\pm =\frac{1}{2}(1\pm ie_7);\quad v^{(n)}_\pm =\frac{1}{2}
e_n(1\pm ie_7)
\end{equation}
it is known~\cite{fourteen} that, in so far they select in octonion space
the preferred direction $e_7$, they define an $SU(3)$ invariant
algebra (see A2), for which $v^{(n)}_+$ and $v^{(n)}_-$ transform
as the $(3)$ and $(\bar 3)$ representations of $SU(3)$
respectively, while $u_+$ and $u_-$ as singlets. Furthermore
$u_+,v^{(n)}_+$ and $u_-,v^{(n)}_-$ define quaternion algebras,
and, following the definitions (11.4) by which $e_0:=1_4$, they
may be concisely indicated with $v^{(m)}_\pm$ for $m=1,2,3,0$ (since
$v^{(0)}_\pm\equiv u_\pm$).

In $\Cl (1,7)$ they become:
\begin{equation}
u_\pm =\frac{1}{2}(1\pm G_9);\quad v^{(n)}_\pm = \frac{1}{2}
G_{4+n}(1\pm G_9)
\end{equation}
and in $\Cl (1,9)$:
$$
U_\pm =\frac{1}{2}(1\pm \cg_{11});\quad V^{(n)}_\pm = \frac{1}{2}
\cg_{6+n}(1\pm \cg_{11})\ , \eqno(11.14')
$$
again concisely expressible as $V^{(m)}_\pm$, with $m=1,2,3,0$. We see
that they contain both the generators of the Clifford algebras and the
projectors apt to insert them in the above Cartan's equations,
and eq.(11.6) becomes:
$$
(P_\mu G^\mu +\sum^3_{n=1}\ P_{4+n}v^{(n)}_\pm +P_8G_8)
u_\pm\Theta =0\ , \eqno(11.6')
$$
which for $u_+\Theta =N$, doublet of Dirac spinors (remember that
$\G_a=\G^{(0)}_a)$ identifies with eq.(6.5) presenting the $SU(2)$
isospin internal symmetry where clearly only the quaternion subalgebra
$e_n$ is
acting. Eq.(11.10) becomes:
$$
(P_a\cg^a+\sum^3_{n=1}\ P_{6+n}V^{(n)}_\pm +P_{10}\cg_{10})
v_\pm\Phi =0\ . \eqno(11.10')
$$

Identifying the reduced spinor $v_+\Phi$ with $\Theta_B$ of
eq.(10.1):
$$
\Theta_B=\begin{pmatrix} b_1\cr b_2\cr b_3\cr b_4\end{pmatrix} =
\begin{pmatrix} B_1\cr B_2\end{pmatrix}\ ,
$$
eq.(11.10$'$) presents, (for $G_A=G^{(0)}_A)$,
an internal symmetry $SU(2)$ acting on the fermion doublets $B_1$
and $B_2$ and another $SU(2)$ is present in the terms
$P_5\cg_5+P_6\cg_6+P_7\cg_7$ acting on the fermions $b_1,b_2$ and
$b_3,b_4$ (for $\G_a=\G^{(0)}_a$).

Here an internal symmetry $SU(3)$ may emerge, and it will be
flavor, since it has isospin $SU(2)$ as a sub-symmetry in the
reduced space. In fact observe that $V^{(n)}_\pm$ for $n=1,2,3$
(saturated by $P_{6+n}$) transform as $(3)$ of $SU(3)$. But it
should also be possible to obtain the known Gell-Mann
representation of the pseudo-octonion algebra~\cite{fifteen} by acting
with the $3$ operators corresponding to $V^{(n)}_\pm$ on
Cartan-standard spinors~\cite{one}, or equivalently, on the vacuum of a
Fock representation of spinor space as in ref.\cite{four}, to obtain, as
minimal left ideals, $3$ spinors representing quarks, as will be
discussed elsewhere. In this way the first two sums in eq.(10.3)
could be realized.

\subsection{$SU(3)$ color}

But there is another $SU(3)$ invariant algebra, contained in $\Cl
(1,7)$ and $\Cl (1,9)$. In fact suppose that, as discussed in
Chapter 7, a local transformation, generated by $G_7G_8$, gives
rise to a gauge potential then eq.(11.6) for $\Cl (1,7)$ may be
written in four equivalent forms as eq.(6.15) in section 6.3:
$$
\left[ (P^\mu -A^\mu_{(m)})G^{(m)}_\mu + \sum^8_{j=5}\
P_jG^{(m)}_j\right] (1\pm G_9)\Theta^{(m)}= 0,\quad m=1,2,3,0
\eqno(11.6'')
$$
where
\begin{equation}
A^{(m)}_\mu = \widetilde{\Theta}^{(m)} G^{(m)}_\mu \Theta^{(m)}
\end{equation}
and $G^{(m)}_A$ are the generators of $\Cl (1,7)$, with
$\Theta^{(m)}$ the corresponding spinors as given in A1,
eqs.(A1.10) to which the arguments of section 6.3 may be applied.
Take now $m=2$:
\begin{equation}
G^{(2)}_\mu = \s_2\otimes\G_\mu
\end{equation}
and substitute $\G_\mu$ with the 3 generators given in (A1.7):
$\G^{(n)}_\mu =\s_n\otimes\g_\mu$, where $n=1,2,3$, then define:
\begin{eqnarray}
-iG^{(2,n)}_\mu &=&\begin{pmatrix} 0&-\s_n\cr \s_n&0\end{pmatrix}
\otimes\g_\mu := e_n\otimes\g_\mu\nonumber \\
iG^{(2)}_9 &=&i\s_3\otimes 1_8=e_7\otimes 1_4\\
G^{(2,n)}_\mu G^{(2)}_9 &:=& e_ne_7\otimes\g_\mu = \hat e_n\otimes
\g_\mu \nonumber
\end{eqnarray}
and with them define the $SU(3)$ invariant complex octonion algebra:
\begin{equation}
u_\pm =\frac{1}{2} (1\pm G_9);\ \  v^{(m)}_{\mu\pm} =
\frac{1}{2}\ e_m\otimes\g_\mu (1\pm G_9) .
\end{equation}

Observe that for $m=0, e_0=1_4$, and then we may interpret it as
representing $G^{(2,0)}_\mu$ by which the spinor $u_+\Theta$ is a
doublet of Dirac spinors and $v^{(0)}_{\mu\pm}$ as well as $u_\pm$
transform as a singlet for $SU(3)$.

We may now introduce these in eq.(11.6$''$) which, taking into
account of (11.6$'$), will become:
$$
\left[ (P^\mu+iA^\mu_{(2,m)})v^{(m)}_{\mu\pm}+\sum^3_{n=1}\
P_{4+n}v^{(n)}_\pm+P_8G_8\right] u_\pm\Theta^{(2,m)}=0\ .
\eqno(11.6''')
$$
These equations identify with those discussed in Chapter 6 and
therefore contain both the $SU(2)$ isospin internal symmetry and
the $SU(2)_L$ one of the electroweak model (in the dynamical
terms).

In a similar way from eq.(11.10) for $\Cl (1,9)$ we obtain:
$$
\left[ (P^\mu -A^\mu_{(m)})\cg^{(m)}_\mu + \sum^{10}_{j=5}\
P_j\cg^{(m)}_j\right] (1\pm \cg_{11})\Phi^{(m)}= 0\ \ \ m=1,2,3,0
\eqno(11.10'')
$$
and we will take:
\begin{eqnarray}
-i\cg^{(2,n)}_\mu &=& \begin{pmatrix} 0&-G^{(n)}_\mu\cr G^{(n)}_\mu
&0\end{pmatrix} := e_n\otimes\G_\mu  \nonumber \\
i\cg^{(2)}_{11} &=& e_7\otimes 1\\
\cg^{(2,n)}_\mu \cg^{(2)}_{11} &=& \hat e_n\otimes \G_\mu \nonumber
\end{eqnarray}
with the $SU(3)$ invariant algebra
\begin{equation}
U_\pm =\frac{1}{2} (1\pm\cg_{11});\ \
V^{(m)}_{\mu\pm}=\frac{-i}{2} \cg^{(2,m)}_{\mu\pm}(1\pm\cg_{11}) =
\frac{1}{2} e_m\otimes \G_\mu (1\pm \cg_{11})
\end{equation}

We may then introduce in Cartan's equation (11.10$''$) the $SU(3)$
invariant algebras and it becomes:
$$
\Biggl[ (P^\mu +iA^\mu_{(2,m)})V^{(m)}_{\mu\pm} +
P_5\cg^{(2,m)}_5+P_6\cg^{(2,m)}_6\Biggr.\hspace{2cm}$$
$$\left.\qquad +\sum^3_{n=1}\ P_{6+n}V^{(n)}_\pm
+P_{10}\cg^{(2,m)}_{10}\right] U_\pm\Phi^{(2,m)}= 0 .
\eqno(11.10''')
$$

Observe that eq.(11.10$'''$), after a dimensional reduction
reduces to eq.(11.6$'''$) above containing $SU(2)$ isospin and
$SU(2)_L$ electroweak, therefore if, as discussed above (for
eq.(11.10$'$)) the $SU(3)$ acting on $U_+=V^{(0)}_+$ and
$V^{(n)}_+$ represents flavour, the one acting on $V^{(0)}_{\mu
+}$ and $V^{(n)}_{\mu +}$ could represent color, for which
$A^{(2,n)}_\mu$ with $n=1,2,3$ could represent the colored gluons.
Observe that, in this interpretation the 3 colors are
characterized by $e_n$; that is by the quaternion subalgebra of
octonions, that is by their 3 imaginary units (Pauli matrices).

In the dimensional reduced (11.6$'''$) $V^\mu_{\mu +}$ identifies
with $v^\mu_+$ giving rise to the electroweak $SU(2)_L$ which
would then appear as correlated with $SU(3)$-color in a parallel
way as $SU(2)$ isospin may be considered as a subgroup of
$SU(3)$-flavour. However, $SU(2)_L$ is not a subgroup since it
does not refer to the same interactions (strong) of $SU(3)$-color.

Also for $SU(3)$-color the arguments on the
possibility of deriving a $3\times 3$ representations of the
corresponding pseudo-octonion algebras may be adopted.

Observe that the proposed geometrical interpretation of color
could give a geometrical explanation on why only uncolored
fermions may appear as free particles; that is, on why only
$SU(3)$-color singlets, are observable. In fact $V^{(0)}_{\mu
+}=1\otimes\G_\mu (1+\cg_{11})$ are $SU(3)$ singlets and
correspond to observable multiplets of Dirac spinors (possibly
trilinear in quarks); while the colored triplets  $V^{(n)}_{\mu
+}$ correspond also to Weyl multiplets which are unobservable
since they obey coupled Dirac equations and transform in each
other for space-time reflections. This is also evident for the
$SU(2)_L$ subgroups of the electroweak model where the chirally
projected electron: $e_L=\frac{1}{2} (1+\g_5)e$ (or proton or
neutron) cannot be observed as a free fermion. There is also a
correlation of the $A_{(2,j)}^\mu -SU(3)$ (with $j=1,2,3)$ gluons
with the $W^\pm_\mu$ and $Z^0_\mu -SU(2)_L$-mesons to be further
analysed, the former bearing the strong charge while the latter
(bilinear in twistors~\cite{eight}) the electroweak one.

In all this the quaternion subalgebra $1$, $e_1, e_2, e_3$, of the
octonion algebra plays a basic role, as it played in the
determination of families in Chapter 10. Therefore quaternions
seem to be the unique origin of isospin, electroweak, 3 families,
and 3 colors.

As mentioned in the Introduction our spinor approach was at first
proposed as an alternative to the traditional one in space-time,
which is also postulated to be ten dimensional and lorentzian, where
however the concept of geometrical point-event is substituted by that of
strings or superstrings, which also allowed the extension of the
theory to the possibly quantizable gravitational field. The two
approaches might not however be as alternative as they might
appear at first sight, and not only because we arrived, through
spinor-geometrical arguments, at the same dimension  and signature, even
if in momentum space, but specially because also strings and
perhaps even superstrings might naturally originate (bilinearly)
from simple or pure spinors as we will briefly show.

\section{STRINGS FROM SPINORS}

The central role which null-vectors and null-lines (lines with
null tangent) played in the last two centuries in the development
of geometry, in the frame of complex analysis, emerged from the
well-known Enneper-Weierstrass parametrization of minimal surfaces
in $\R^3$, in the form:
\begin{equation}
X_j(u,v)=X_j(0,0)+{\rm Re} \int^{u+iv}_c Z_j(\alpha ) d\alpha
;\quad j=1,2,3
\end{equation}
where $X_j(u,v)$ are the orthonormal coordinates of the points of
a surface, which is minimal provided $Z_j(\alpha )$ are the
holomorphic coordinates of a null $\C^3$-vector, and $c$ is any
path in the complex plane $(u,v)$ starting from the origin. The
correlation with spinors associated with null $\C^3$-vectors, are
given by eqs.(3.1$'$), where $Z_j=\frac{1}{2}\langle
\varphi^t\epsilon \s_j\varphi\rangle$ satisfy identically
eq.(3.2). It was shown in ref.\cite{sixteen} that, by considering
$\varphi$ as a Weyl spinor associated with $\Cl(3,1)$ eq.(12.1)
may easily be extended to $\R^{3,1}$. For a Majorana spinor
associated with $\Cl(3,1)=R(4)$ the corresponding equation gives
the representation of a string in the form:
\begin{equation}
X_\mu(\s,\tau )=X_\mu(0,0)+ \int^{\s+\tau}_0 t^+_\mu (\alpha )
d\alpha \int^{\s-\tau}_0 t^-_\mu (\beta ) d\beta ;\ \ \ \mu
=0,1,2,3
\end{equation}
where $t^\pm_\mu$ are real, null vectors bilinearly constructed in
terms of Majorana spinors. It was further shown~\cite{seventeen} that the
above formalism may be extended to higher dimensional Clifford
algebras and corresponding spinor spaces and that, whenever real
Majorana spinors are admitted, strings will be naturally obtained
as integrals of bilinear null vectors in terms of them, which
applies in particular to the case of $\Cl (9,1)$, in which frame
the string approach to the gravitational field is notoriously
contained. Propositions 1 and 2 of the present paper may be
adopted also in the case of strings such that imbedding simple
spinor spaces in higher dimensional ones implies the corresponding
imbedding of strings.

In this way the approach presented in this paper could not only be
compatible with, but even be at the origin of string theory, in
accordance with the Cartan's conception of simple spinor geometry
on which this paper is based. In fact Cartan conjectured that the
fundamental geometry of nature be that of simple spinors, out of
which euclidean geometry derives, in so far simple spinors
bilinearly generate null vectors and sums of null vectors may give
the ordinary non null vectors of euclidean geometry~\cite{one}.
Now the integrals above, defining strings, may be conceived as
continuous sums of null vectors and therefore, in the spirit of
Cartan, they could well be thought as the intermediate step
between simple spinor- and euclidean-geometry where then they
could well have to substitute the purely euclidean concept of
point-event.

A bilinear parametrization of covariant strings and superstrings
theories in terms of Majorana-Weyl spinors associated with
$\Cl(1,9)$ was proposed in refs.\cite{eighteen}, \cite{nineteen}. The
general solutions of the equations of motion are obtained through the use
of octonions, which might render transparent, through the triality
automorphism, also the geometrical origin of supersymmetry.
Recently, pure spinors have been successfully adopted in
superstring theory~\cite{twenty}.

\section{FURTHER GEOMETRICAL ASPECTS AND CONCLUDING REMARKS}

We attempted to show how the elegant geometry of simple or pure
spinors could be helpful for throwing some light on several
aspects of the tantalizing world of the elementary constituents of
matter and, if correlated with the division algebras, naturally
emerging from that geometry, could plainly explain the origin of
charges, internal symmetry groups and families. The strict
correlation of division algebras with elementary particle physics
has been also shown by G.M. Dixon~\cite{twentyone}.

In this preliminary approach we concentrated our attention on the
main features of Cartan's conception of simple spinors as
isomorphic (up to a sign) to maximal totally null planes defining
projective quadrics and corresponding compact manifolds (spheres)
all imbedded one in the other up to that in a ten dimensional
lorentzian momentum space. The basic role of the geometry of
simple or pure spinors is manifested not only by the fact that it
allows the simultaneous embeddings of spinor spaces and of the
corresponding null vector spaces, but also by the fact that the
equations of motions, originally in momentum spaces, may be
written soley in spinor spaces, and the necessary condition for
the existence of solutions $(p_ap^a=0)$ become identities, if, as
suggested by Cartan, the components of momenta are expressed
bilinearly in terms of spinors, if these are simple or pure; and
also the equations of motion may become identities in spinor
spaces. That is, they are true in the whole simple or pure spinor
space independently from momentum space or space-time. This could
then assign a purely geometrical, that is spinorial, origin to the
equations of motion; and then to the laws they represent, which
are then to be conceived, so to say, outside space-time: in the
realm of simple or pure spinor-geometry. Obviously if a
$p_\mu$-dependent (or $x_\mu$-dependent) spinor-solution is
inserted in the equations of motion they also become identities,
however in the whole momentum space; meaning the validity for any
$p$ or, equivalently, the validity of the evolution of the
phenomena contemplated by the laws, for the whole space-time.
Maxwell's equations are an example: the electromagnetic tensor
$F^\pm_{\mu\nu}$ derives, bilinearly, from the Weyl's spinors
$\varphi_+=\frac{1}{2} (1+\g_5)\psi$ and $\varphi_-=\frac{1}{2}
(1-\g_5)\psi$ (see eq.(3.13)), which determine then its properties
(as noted already by Cartan~\cite{one}), while their solutions:
$F^\pm_{\mu\nu}(x,t)$ well represent propagation of light both
here, now, and on the distant galaxies as they were bilions of
years ago. The same is true for the Cartan's
equations\footnote{Only Cartan's equations for simple or pure
spinors appear to present this double possibility of becoming
identities: both in spinor space (if momenta are bilinear
functions of spinors) and in momentum space or space-time (if
spinors are functions of momenta or space-time; that is
solutions). This double possibility allows an epistemological
correlation between the laws of physics (the equations) and the
consequent phenomena (their solutions): the former to be conceived
outside space-time ``are'' in Parmenides conception of ``to be''
which ``neither was nor will be but always is''. The phenomena
instead universally evolve in space-time, as foreseen by the
solutions of those equations or laws, and then ``become'' well
representing Eraclito's ``panta rei''. Cartan's simple spinor
geometry could then well represent an example of synthesis between
these two apparently antithetic philosophies.} in this paper where
momenta $p_\mu$ (or space-time coordinates $x_\mu$, that is
$t^\pm_\mu$ in eq.(12.2)) are bilinear functions of spinors and as
such make identities both the necessary conditions for the
existence  of solutions and the equations of motion, while their
solutions are spinors, functions of $x_\mu$, making them
identities in space-time. Then simple spinors determine bilinearly
$p_\mu$ (or $x_\mu$) while Cartan's equations represent the
physical laws, and their solutions,  the universal evolution of
the phenomena. In this way the origin of the physical laws ruling
the behaviour of the elementary  constituents of matter: the
fermions\footnote{Bosons, in this frame, result bi- or
multi-linear in spinors and represent either composite states,
with inner structure, as notoriously happens in nuclear physics
(helium) or just new euclidean scalar, vector or tensor fields,
without inner structure, as in the case of the electromagnetic
gauge potential and, presumably, of the colored gauge gluons.},
might naturally lay in simple spinor geometry, which, in turn,
following Cartan, may be conceived as the elementary  constituent
of euclidean geometry.

The geometry of spinors we dealt with is however very rich and
presents further aspects like the constraint equations for simple
spinors and the triality symmetry for eight dimensional spaces,
which, while geometrically fundamental, could also be of interest
for comparing the mathematical results with the physical world.
Specially, the constraint-equations, correlated with simplicity,
and then allowing the whole construction, through Propositions 1
and 2, might play important roles (as recent results seem to
indicate~\cite{twenty}) and could also furnish interesting
indications on the  stability of the proton \cite{eight} and on
the possibility of theoretical predictions, which were not dealt
with in the present preliminary paper, to be completed with a
study of the physical meaning of the steadily increasing values of
the invariant masses characterizing the spheres imbedded in each
other, presumably  accompanied by the increasing values of the
corresponding charges, which also should be geometrically
determined as will be discussed elsewhere.

Should this approach have further confirmations in the physical
world, then the old Cartan's conjecture on the fundamental role of
simple spinor geometry, at the origin of euclidean geometry, would
not only obtain a strong support, but it could also be extended to
the origin of the geometry of quantum mechanics, in so far, the
Cartan's equations defining simple spinors, may be interpreted as
equations  of motion for fermions, the most elementary
constituents of matter, and they are quantum equations (in first
quantization, up to the definition of Planck's constant). A
remarkable aspect of these equations is that, at first, they
naturally appear in momentum space, sometimes called the space of
velocities; after all a natural space for the description of
elementary motions. If furthermore, as it would here appear, in
momentum space several of the mysterious aspects of the elementary
constituents of matter, might be easily explained and well
understood, in purely geometrical (spinorial) form, one could also
hope that in this space (of velocities or rather of elementary
motions) several of the difficulties, which notoriously hinder the
understanding of quantum mechanics, when described in space-time,
could be substantially mildered if not disappear.

One could then be induced to conjecture that while space-time
remains the ideal arena for the euclidean description of the
classical form of mechanics (of astronomy), it is momentum space
the ideal one for the description, by means of the elementary form
of geometry: the Cartan's one of simple spinors, of the elementary
form of mechanics: quantum mechanics (of fermions); the two being
possibly correlated by conformal inversions or
reflections~\cite{twentytwo}.

\section*{Acknowledgments}

The author expresses his gratitude to L. Bonora, G. Calucci, L.
Dabrowski,\break 
T. Dray, B. Dubrovin, P. Furlan, R. Jengo, G. Landi, S.
Majid, A. Masiero, C. Monogue, P. Nurowski, N. Paver, S.
Randjbar-Daemi, C. Reina and A. Trautmant for helpful discussions
and suggestions.

\newpage

\appendix

\renewcommand{\thesection}{A\arabic{section}}

\centerline{\Large {\bf APPENDICES}}
\vspace{1cm}

\section{CLIFFORD ALGEBRAS, SPINORS, OPTICAL VECTORS, CARTAN'S
EQUATIONS: FROM $\Cl (1,3)$ TO $\Cl (1,9)$}

\section*{Notations}

Indices:
\begin{equation}
n=1,2,3;\ \ \mu =1,2,3,0;\ \ a=\mu ,5,6;\ \ A=a,7,8;\ \ \alpha
=A,9,10 . \
\end{equation}

\noindent Spinors:
\begin{equation}
\begin{array}{ll}
\psi\in S_4:\Cl (1,3)\ =\End\ S_4; &\tilde\psi :=\psi^\dagger\gamma_0,\\
\Psi\in S_8:\Cl (1,5)\ =\End\ S_8; &\tilde\Psi :=\Psi^\dagger\Gamma_0,\\
\Theta\in S_{16}: \Cl (1,7)=\End\ S_{16}; &\tilde\Theta :=\Theta^\dagger G_0,\\
\Phi\in S_{32}:\Cl (1,9)=\End\ S_{32}; &\tilde\Phi
:=\Phi^\dagger\cg_0.
\end{array}
\end{equation}

\section*{$\Cl (1,3)$; {\normalsize generators $\g_\mu$:}}
\begin{equation}
\g_\mu :\ \ \g_n=\begin{pmatrix} 0&-\s_n\cr \s_n&0\end{pmatrix}
=-i\s_2\otimes \s_n;\ \ \g_0=\begin{pmatrix} 0&1\cr
1&0\end{pmatrix} =\s_1\otimes 1_2;
\end{equation}
volume element
\begin{equation}
\g_5=\begin{pmatrix} 1&0\cr 0&-1
 \end{pmatrix} =\s_3\otimes 1_2.
\end{equation}
Optical vectors:
\begin{equation}
p^\pm_\mu = \tilde\psi (1\pm\g_5)\g_\mu\psi ,
\end{equation}
(the factor $\frac{1}{4}$ was eliminated for brevity: here and in
the following we are dealing with null vectors which may anyhow be
arbitrarily scaled.)

\noindent Cartan's equation:
\begin{equation}
p^\pm_\mu\g^\mu (1\pm\g_5)\psi =0
\end{equation}
\section*{$\Cl (1,5)$; {\normalsize  generators $\Gamma_a$:}}
\begin{equation}
\begin{array}{rl}
\text{Weyl:} &\left\{\begin{array}{rl} \Gamma^{(1)}_\mu
=\s_1\otimes \g_\mu; &\Gamma^{(1)}_5 =-i\s_1\otimes \g_5,
\Gamma^{(1)}_6 =-i\s_2\otimes 1_4, \Gamma^{(1)}_7 =\s_3\otimes
1_4\\
\Gamma^{(2)}_\mu =\s_2\otimes \g_\mu; &\Gamma^{(2)}_5
=-i\s_2\otimes \g_5, \Gamma^{(2)}_6 =-i\s_1\otimes 1_4,
\Gamma^{(2)}_7 =\s_3\otimes 1_4\end{array}\right.\\
\text{Pauli:}&\quad\  \Gamma^{(3)}_\mu =\s_3\otimes \g_\mu;\ \
\Gamma^{(3)}_5 =-i\s_1\otimes 1_4, \Gamma^{(3)}_6 =-i\s_2\otimes
1_4, \Gamma^{(3)}_7 =\s_3\otimes \g_5\\
\text{Dirac:} &\quad\  \Gamma^{(0)}_\mu =1_2\otimes \g_\mu; \ \
\Gamma^{(0)}_5 =-i\s_1\otimes \g_5, \Gamma^{(0)}_6 =-i\s_2\otimes
\g_5, \Gamma^{(0)}_7 =\s_3\otimes \g_5
\end{array}
\end{equation}
Optical vectors (Corollary 2):
\begin{equation}
p^\pm_a:p^\pm_\mu =\tilde\psi\g_\mu\psi;\ \
p^\pm_5=-i\tilde\psi\g_5\psi,\ p^\pm_6=\pm\tilde\psi\psi
\end{equation}
or:
$$p^\pm_a=\tilde\Psi \Gamma_a(1\pm\Gamma_7)\Psi
\eqno({\rm A}1.8')$$
Cartan's equation:
\begin{equation}
p^\pm_a\Gamma^a(1\pm\Gamma_7)\Psi =0
\end{equation}
\section*{$\Cl (1,7)$; {\normalsize  generators $G_A$:}}
\begin{equation}
\begin{array}{rl}
\text{Weyl:} &\left\{\begin{array}{rl} G^{(1)}_a =\s_1\otimes
\Gamma_a; &G^{(1)}_7 =-i\s_1\otimes \Gamma_7, G^{(1)}_8
=-i\s_2\otimes 1_8, G^{(1)}_9 =\s_3\otimes
1_8\\
G^{(2)}_a =\s_2\otimes \Gamma_a; &G^{(2)}_7 =-i\s_2\otimes
\Gamma_7, G^{(2)}_8 =-i\s_1\otimes 1_8, G^{(2)}_9 =\s_3\otimes
1_8\end{array}\right.\\
\text{Pauli:}&\quad\  G^{(3)}_a =\s_3\otimes \Gamma_a;\ \
G^{(3)}_7 =-i\s_1\otimes 1_8, G^{(3)}_8 =-i\s_2\otimes 1_8,
G^{(3)}_9
=\s_3\otimes \Gamma_7\\
\text{Dirac:} &\quad\  G^{(0)}_a =1_2\otimes \Gamma_a;\ \ G^{(0)}_7
=-i\s_1\otimes \Gamma_7, G^{(0)}_8 =-i\s_2\otimes \Gamma_7,
G^{(0)}_9 =\s_3\otimes \Gamma_7
\end{array}
\end{equation}
Optical vectors (Corollary 2):
\begin{equation}
P^\pm_A:P^\pm_a =\tilde\Psi\Gamma_a\Psi;\ \
P^\pm_7=-i\tilde\Psi\Gamma_7\Psi,\ P^\pm_8=\pm\tilde\Psi\Psi
\end{equation}
or:
$$P^\pm_A=\tilde\Theta G_A (1\pm G_9)\Theta
\eqno({\rm A}1.11')$$
Cartan's equation:
\begin{equation}
P^\pm_AG^A(1\pm G_9)\Theta =0 .
\end{equation}
If
\begin{equation}
\Theta = \begin{pmatrix} \Psi_+\cr \Psi_-\end{pmatrix}
\end{equation}
Cartan's equation:
$$
(P_a\Gamma^a-iP_7\Gamma^7\pm P_8)\Psi_\pm =0 . \eqno({\rm
A}1.12')$$

Taking for $\Gamma_a$ the Dirac's rep. $\Gamma^0_a$ as in (A1.7)
we obtain:\\
Cartan's equation:
$$
(P_\mu\g^\mu -i\vec\pi\cdot\vec\s\otimes\g_5\pm P_8)\Psi_\pm =0
\eqno({\rm A}1.12'')$$
where $\pi_{1,2,3}$ are pseudo scalars of
$\R^{3,1}$ (see eq.(6.6)), and, therefore
\begin{equation}
\pi_{1,2,3}\otimes \g_5=P_{5,6,7}
\end{equation}
are scalars of $\R^{3,1}$; we will indicate them with
$P_{5,6,7}:=s_{4+n}$, with $n=1,2,3$.

If we now represent with $q_n=-i\s_n$ the quaternion units we
obtain the Cartan's equation (A1.12$''$) in the quaternionic form:
$$
(P_\mu\g^\mu +s_{4+n}q^n\pm P_8)\Psi_\pm =0 \eqno({\rm
A}1.12''')$$
\section*{$\Cl (1,9)$; {\normalsize generators
$\cg_\alpha$:}}
\begin{equation}
\begin{array}{rl}
\text{Weyl:} &\left\{\begin{array}{rl} \cg^{(1)}_A =\s_1\otimes
G_A; &\cg^{(1)}_9 =-i\s_1\otimes G_9, \cg^{(1)}_{10}
=-i\s_2\otimes 1_{16}, \cg^{(1)}_{11} =\s_3\otimes
1_{16}\\
\cg^{(2)}_A =\s_2\otimes G_A; &\cg^{(2)}_9 =-i\s_2\otimes G_9,
\cg^{(2)}_{10} =-i\s_1\otimes 1_{16}, \cg^{(2)}_{11} =\s_3\otimes
1_{16}\end{array}\right.\\
\text{Pauli:}&\quad\  \cg^{(3)}_A =\s_3\otimes G_A;\ \ \cg^{(3)}_9
=-i\s_1\otimes 1_{16}, \cg^{(3)}_{10} =-i\s_2\otimes 1_{16},
\cg^{(3)}_{11} =\s_3\otimes
G_9\\
\text{Dirac:} &\quad\  \cg^{(0)}_A =1_2\otimes G_A;\ \
\cg^{(0)}_9 =-i\s_1\otimes G_9, \cg^{(0)}_{10} =-i\s_2\otimes G_9,
\cg^{(0)}_{11} =\s_3\otimes G_9
\end{array}
\end{equation}
Optical vectors (Corollary 2):
\begin{equation}
P^\pm_A:\tilde\Theta G_A\Theta ;\ \ P^\pm_9=-i\tilde\Theta
G_9\Theta ;\ \ P^\pm_{10} = \pm\tilde\Theta\Theta ,
\end{equation}
or:
$$P^\pm_\alpha =\tilde\Phi \cg_\alpha (1\pm \cg_{11})\Phi
\eqno({\rm A}1.16')$$
Cartan's equation:
\begin{equation}
P_\alpha\cg^\alpha (1\pm\cg_{11})\Phi = 0 .
\end{equation}
If
\begin{equation}
\Phi = \begin{pmatrix} \Theta_+\cr \Theta_-\end{pmatrix}
\end{equation}
Cartan's equation (for the representation $G^{(1)}_A$ from
(A1.10)):
$$
(P_AG^A-iP_9G^9\pm P_{10})\Theta_\pm =0=\begin{pmatrix} \pm
P_{10}-iP_9&P_a\Gamma^a-iP_7\Gamma^7-P_8\cr
P_a\Gamma^a-iP_7\Gamma^7+P_8 &\pm P_{10}+iP_9\end{pmatrix}
\Theta_\pm \eqno({\rm A}1.18')
$$
The non diagonal terms are of the form (A1.12$'$). Setting them in the form
(A1.12$''$), and taking into account of (A1.4) we have that, in (A1.18$'$),
after setting $\Gamma_a=\Gamma^{(0)}_a$:
\begin{equation}
G^{(1)}_{5,6,7}(\Gamma^{(0)}_a) :=G^{(1,0)}_{5,6,7} = \begin{pmatrix}
0&-i\s_{1,2,3}\cr -i\s_{1,2,3}&0\end{pmatrix} \otimes
\g_5=-i\s_1\otimes\s_{1,2,3}\otimes\g_5
\end{equation}
If we impose that $P_{5,6,7}$ are scalars of $\R^{3,1}$ the $\g_5$
factor becomes $1$: see eq.(A1.14).

For the representation $G^{(2)}_A$ from (A1.10) instead, Cartan's equation
becomes:
$$
\begin{pmatrix}
\pm P_{10}-iP_9 & -iP_a\G^a-P_7\G^7 +iP_8\cr -iP_a\G^a+P_7\G^7
+iP_8 &\pm P_{10}+P_9\end{pmatrix} \Theta_\pm =0 \eqno({\rm
A}1.18'')
$$
With the same procedure as before one obtains that, this time:
\begin{equation}
G^{(2)}_{5,6,7}(\G^{(0)}_a):= G^{(2,0)}_{5,6,7} = \begin{pmatrix}
0 &-\s_{1,2,3}\cr \s_{1,2,3} &0\end{pmatrix} \otimes \g_5
=-i\s_2\otimes\sigma_{1,2,3}\otimes \g_5
\end{equation}
Again $\g_5=1$, for $P_{5,6,7}$ scalars.\\
According to Corollary 2 applied to $\Cl (1,9)$ we have that the vector with
components:
\begin{equation}
\cp^\pm_\alpha =\tilde\Phi\cg_\alpha\Phi  ;\ \
\cp^\pm_{11}=-i\tilde\Phi\cg_{11}\Phi ;\ \ \cp^\pm_{12}=\pm
\tilde\Phi\Phi ,
\end{equation}
is null in $\R^{1,11}$, for $\Phi$ simple Weyl of $\Cl_0 (1,11)$.
They give rise to Cartan's equation:
\begin{equation}
(\cp_\a\cg^\a -i\cp_{11}\cg^{11}\pm\cp_{12})\phi_\pm = 0 =
\begin{pmatrix}
\pm\cp_{12}-i\cp_{11} &\cp_AG^A-i\cp_9G^9-\cp_{10}\cr
\cp_AG^A-i\cp_9G^9+\cp_{10} &\pm\cp_{12}+i\cp_{11}
\end{pmatrix}
\end{equation}
where the representation $\cg^{(1)}_\a$ was taken from (A1.15). If
we now insert in it $G^{(0)}_A$ from (A1.10) we obtain:
\begin{equation}
\cg^{(1)}_{7,8,9}(G^{(0)}_A):=\cg^{(1,0)}_{7,8,9}=
-i\s_1\otimes\s_{1,2,3}\otimes\G_7 .
\end{equation}

For the choice $\cg^{(2)}_\a$ instead we would have obtained:
\begin{equation}
\cg^{(2)}_{7,8,9}(G^{(0)}_A):=\cg^{(2,0)}_{7,8,9}=
-i\s_2\otimes\s_{1,2,3}\otimes\G_7 .
\end{equation}

\section{MATRIX REPRESENTATIONS OF OCTONIONS}

\section*{Notations}

\noindent Octonions:
\begin{equation}
\begin{array}{rl}
{\bf o} &=e_0p_8+\mathop{\sum}\limits^7_{j=1}\ p_je_j\\
\bar{\bf o} &=e_0p_8-\mathop{\sum}\limits^7_{j=1}\ p_je_j\\
{\bf o}\bar{\bf o} &=\bar{\bf o}{\bf o}
=\mathop{\sum}\limits^8_{j=1}\ p^2_j
\end{array}
\end{equation}
where $p_j\in\R$, $e_0$ is the identity and $e_1,e_2,\dots e_7$
are the octonion imaginary anticommuting units for which we adopt
the notation:
\begin{equation}
\begin{array}{c}
e_n =\{ e_1,e_2,e_3\};\ \ e_{n+3}=\{e_4,e_5,e_6\} :=\hat e_n=\{
\hat e_1,\hat
e_2,\hat e_3\} ;\\
e_{n+3}=\hat e_n :=e_ne_7=-e_7e_n .
\end{array}
\end{equation}
The octonion algebra $0$:
\begin{equation}
\begin{array}{rl}
e_\ell e_m &= -\delta_{\ell m}e_0+\epsilon_{\ell mn}e_n\\
e_\ell \hat e_m &= -\delta_{\ell m}e_7-\epsilon_{\ell mn}\hat{e}_n\\
\hat e_\ell \hat e_m &= -\delta_{\ell m}e_0-\epsilon_{\ell mn}e_n
\end{array}
\end{equation}
where $\ell ,m,n=1,2,3$.

The quaternion subalgebra $Q$ is:
\begin{equation}
Q=\{ e_0,e_1,e_2,e_3\}
\end{equation}
Define
\begin{equation}
\hat Q :=Qe_7=\{ e_7,\hat e_1,\hat e_2,\hat e_3\}
\end{equation}
The non associative octonion algebra $0$ may be graded as follows:
\begin{equation}
Q=Q\oplus \hat Q
\end{equation}

\section*{Matrix representations}

Anti De Sitter Clifford algebra $\Cl (2,3)$ with generators
\begin{equation}
\g_n=\begin{pmatrix} 0 &-\s_n\cr \s_n &0\end{pmatrix};\ \
\g_0=\begin{pmatrix} 0 &1\cr 1 &0\end{pmatrix}\ \
\g_5=\begin{pmatrix} 1 &0\cr 0 &-1\end{pmatrix}
\end{equation}
Define:
\begin{equation}
e_0:=\begin{pmatrix} 1 &0\cr 0 &1\end{pmatrix} ;\ \
e_n:=\g_n;\ \  e_7:=i\g_5 =
\begin{pmatrix} i &0\cr 0 &-i\end{pmatrix}
\end{equation}
\begin{equation}
e_ne_7:=e_{n+3}:=\hat e_n :=\begin{pmatrix} 0 &i\s_n\cr i\s_n &0\end{pmatrix}
\end{equation}
they close the octonion algebra either by modifying the
multiplication rules of matrices~\cite{thirteen} or by defining the
product
$\odot$ as follows:
\begin{equation}
e_j\odot e_k:=\delta_{jk}e_je_k+(1-\delta_{jk})i\g_5\g_0e_j\g_0e_k\g_0
\end{equation}
(on the r.h.s. products are ordinary matrix multiplications) for
$j,k=1,2,\dots 6$. While for $e_7$ the multiplication is defined
by (A2.9) and obviously by $\hat e_ne_7=-e_n$.

The products $\odot$ and (A2.9) close the octonion algebra of the imaginary
units $e_0,e_1,\dots e_7$.

Another matrix representation of the octonion units may be
obtained by defining:
$$
e_0:=\begin{pmatrix} 1 &0\cr 0 &1\end{pmatrix} ;\ \  e_n:=
\begin{pmatrix} 0 &-i\s_n\cr -i\s_n &0\end{pmatrix} ;\ \
e_7:= \begin{pmatrix} i &0\cr 0 &-i\end{pmatrix} ;
\eqno({\rm A}2.8')
$$
$$
e_ne_7:=e_{n+3}:=\hat e_n=\begin{pmatrix} 0 &-\s_n\cr \s_n
&0\end{pmatrix} \eqno({\rm A}2.9')
$$
after which the $\odot$ product becomes
$$
e_j\odot e_n =\delta_{jk}e_je_k+(1-\delta_{jk})\g_0e_j\g_0e_k\g_0
\eqno({\rm A}2.10')
$$
For $j,k=1,2,\dots 6$, where $\g_0=\begin{pmatrix} 0 &1\cr 1 &0\end{pmatrix}$.
The rest as before.

The associator between octonions is defined as usual with respect
to the $\odot$ products; in particular the one between the
imaginary units $e_j$ is defined by
\begin{equation}
(e_i,e_j,e_k):=(e_i\odot e_j)\odot e_k -e_i\odot (e_j\odot e_k)
\end{equation}
and one easily finds the usual seven non zero ones.\\

\section*{Complex octonions}

are defined by:
\begin{equation}
u_\pm=\frac{1}{2}(1\mp ie_7)=\frac{1}{2}(1\pm\g_5):
v^{(n)}_\pm=\frac{1}{2}(e_n\mp i\hat e_n)=e_n\frac{1}{2}(1\pm\g_5)
\end{equation}
where the equalities with the chiral projectors $\frac{1}{2}(1\pm \g_5)$ are
valid for both the proposed matrix representation of the units $e_j$.

The complex octonions satisfy to the following multiplication rules:
\begin{equation}
\begin{array}{l}
u^2_+=u_+;\ \ u_+u_-=0;\ \ v^{(n)}_+u_+= u_-v^{(n)}_+
\\[1.5ex]
u_+v^{(n)}_+=v^{(n)}_+ u_-=0;\ \ v^{(\ell )}_+
v^{(m)}_-=-\delta_{\ell m}u_0;\ \ v^{(\ell )}_+
v^{(m)}_+=\epsilon_{\ell mn} v^{(n)}_- .
\end{array}
\end{equation}
and the ones obtained substituting $u_+,v^{(n)}_+$ with
$u_-,v^{(n)}_-$ and viceversa. The corresponding algebra is
$SU(3)$ invariant if $(v^{(1)}_+v^{(2)}_+v^{(3)}_+)$ and $(
v^{(1)}_-v^{(2)}_-v^{(3)}_-)$ transform like the $(3)$ and $(\bar
3)$ representations of $SU(3)$ and $u_+,u_-$ like its singlet
representation. This $SU(3)$ is the subgroup of the $G_2$ group of
automorphism of the octonion algebra since $e_7$, that is $i\g_5$
in our case, is
chosen as a preferred direction.\\

\section*{Octonions in $\Cl (1,7) - \Cl (1,9)$ and in the
corresponding Cartan's equations}

\subsection*{a) In interaction terms}

\noindent$\mathbf{\Cl (1,7)}$.\ Consider now eq.(A1.20) for the
generators $G^{(2,0)}_5,G^{(2,0)}_6, G^{(2,0)}_7$ and the
definition (A2.8) of $e_n$ we may write it now in the form
\begin{equation}
G_{4+n} = e_n\otimes\g_5
\end{equation}
Now $G_5,G_6,G_7$ are the generators which, in eqs.(A1.12$''$) and
(A1.12$'''$), give origin to the $SU(2)$ symmetry of isospin. They
may be then, together with the unit $e_0$, be identified with the
quarternionic subalgebra $Q$ of eq.(A2.4) of octonions. In the
frame of $\Cl (1,7)$ we may also identify the generators of $\hat
Q=Qe_7$.

In fact since $G^{(2)}_9=\s_3\otimes 1_8$ (see (A1.10)) we define
$$
iG_9=e_7\otimes 1_4
$$
\begin{equation}
{\rm and}\hspace{10cm}
\end{equation}
$$
iG_{4+n}G_9=e_ne_7\otimes\g_5=\hat e_n\otimes\g_5 .
$$
Again the $\g_5$ factor equals one if $P_{5,6,7}$ are scalars of
$\R^{3,1}$. These seven elements of $\Cl (1,7)$ will then close
the octonion algebra if we define it with the product $\odot$,
defined in (A2.10). If we start from the definition (A1.19) of
$G_{5,6,7}$ and adopt definitions (A2.8$'$) for the octonion units
$e_j$, we obtain again (A2.15) which closes the octonion algebra
however with (A2.10$'$).

Let us now define the complex octonions for $\Cl (1,7)$. They are:
\begin{equation}
u_\pm =\frac{1}{2} (1\pm G_9),\qquad v^{(n)}_\pm =\frac{1}{2}\
G_{4+n} (1\pm G_9) .
\end{equation}
They all contain the projectors $\frac{1}{2} (1\pm G_9)$ which
operate the dimensional reduction; and their algebra is
associative, therefore we may operate with them, normally in
spinor space.\\

\noindent$\mathbf{\Cl (1,9)}$.\ The procedure may be repeated,
starting from eqs.(A1.25) and (A1.26) and we obtain:
\begin{equation}
\begin{array}{rl}
G_{6+n} &=e_n\otimes\G_7,\ \ i\cg_{11}=e_7\otimes 1_8,\\[1.5ex]
&i\cg_{6+n}\cg_{11}=e_ne_7\otimes\G_7=\hat e_n\otimes\G_7 ,
\end{array}
\end{equation}
while the complex octonions are
\begin{equation}
u_\pm =\frac{1}{2} (1\pm\cg_{11}),\qquad v^{(n)}_\pm =\frac{1}{2}\
\cg_{6+n}(1\pm \cg_{11})
\end{equation}

\subsection*{b) In dynamical terms}

We have seen in section 10 that the possible families of baryon-
and lepton-pairs which may be obtained through dimensional
reduction of $\Theta_B$ and $\Theta_\cll$ to fermion doublets are
characterized by the representations of $\G_\mu$.  Precisely
\begin{equation}
\G^{(1,2)}_\mu =\s_{1,2}\otimes\g_\mu ;\ \ \G^{(3)}_\mu
=\s_{3}\otimes\g_\mu ;\ \ \G^{(0)}_\mu =1_2\otimes\g_\mu ,
\end{equation}
characterize Weyl, Pauli, Dirac doublets respectively where $\g_\mu$
were generators of $\Cl (3,1)$. With this convention we may then
adopt the first four generators of $G^{(2)}_\mu$ of $\Cl (1,7)$,
for the first 3 $\G_\mu$ in (A2.19), to be:
\begin{equation}
-G^{(2,n)}_\mu =-i\s_2\otimes\G^{(n)}_\mu =\begin{pmatrix} 0&
-\s_n\cr \s_n &0\end{pmatrix} \otimes\g_\mu := e_n\otimes\g_\mu
\end{equation}
where we have adopted the representation (A2.7) of $e_n$. Defining
now
$$
iG^{(2)}_9 :=i\s_3\otimes 1_8=e_7 \otimes 1_4
$$
and\hfill (A2.20$'$)
$$
G^{(2,n)}_\mu G^{(2,n)}_9:=\hat e_n\otimes \g_\mu
$$
we obtain in $\Cl (1,7)$ another octonion algebra. The complex
octonions will be
\begin{equation}
u_\pm =\frac{1}{2} (1\pm G_9);\qquad v^{(n)}_\pm = G^{(2,n)}_\mu
(1\pm G_9)
\end{equation}
The same algebras may be also easily found in $\Cl (1,9)$ with the
substitution of $\g_\mu$ with $\G_\mu$ and of $G_9$ with
$\cg_{11}$.

\end{document}